\documentclass[twocolumn,english,aps,prb,showpacs,superscriptaddress,floats,amsmath,amssymb,floatfix]{revtex4}

\usepackage[latin9]{inputenc}
\usepackage{color}
\usepackage{amsmath}
\usepackage{graphicx}
\usepackage{graphics}
\usepackage{gensymb}
\usepackage{subfigure}
\usepackage{pdftexcmds}
\usepackage{ifpdf}
\usepackage{amssymb}
\usepackage{dsfont}

\makeatletter
\@ifundefined{textcolor}{}
{%
 \definecolor{BLACK}{gray}{0}
 \definecolor{WHITE}{gray}{1}
 \definecolor{RED}{rgb}{1,0,0}
 \definecolor{GREEN}{rgb}{0,1,0}
 \definecolor{GREEN2}{rgb}{0,0.4,0}
 \definecolor{BLUE}{rgb}{0,0,1}
 \definecolor{CYAN}{cmyk}{1,0,0,0}
 \definecolor{MAGENTA}{cmyk}{0,1,0,0}
 \definecolor{YELLOW}{cmyk}{0,0,1,0}
 \definecolor{YELLOW2}{cmyk}{0,0,1,0.6}
 \definecolor{ORANGE}{rgb}{1,0.22,0}

 }

\def \Sp {\mathbf{S}}
\def \mv {\mathbf{m}}
\def \rv {\mathbf{r}}

\def \dv {\bf{\delta}}

\def \be {\begin{equation}}
\def \ee {\end{equation}}
\def \bea {\begin{eqnarray}}
\def \eea {\end{eqnarray}}
\allowdisplaybreaks
\makeatother

\begin{document}

\title{Stability of skyrmions in perturbed ferromagnetic chiral magnets}

 \author{S.A. Osorio}
 \affiliation{Instituto de F\'isica de L\'iquidos y Sistemas Biol\'ogicos, CCT La Plata, CONICET and Departamento de F\'isica, Facultad de Ciencias Exactas, Universidad Nacional de La Plata, C.C. 67, 1900 La Plata, Argentina}
 \author{M.B. Sturla}
\affiliation{Instituto de F\'isica de L\'iquidos y Sistemas Biol\'ogicos, CCT La Plata, CONICET and Departamento de F\'isica, Facultad de Ciencias Exactas, Universidad Nacional de La Plata, C.C. 67, 1900 La Plata, Argentina}
 \author{H.D. Rosales}
 \affiliation{Instituto de F\'isica de L\'iquidos y Sistemas Biol\'ogicos, CCT La Plata, CONICET and Departamento de F\'isica, Facultad de Ciencias Exactas, Universidad Nacional de La Plata, C.C. 67, 1900 La Plata, Argentina}
\affiliation{Facultad de Ingenier\'ia, Universidad Nacional de La Plata, 1900 La Plata, Argentina}
 \author{D.C. Cabra}
 \affiliation{Instituto de F\'isica de L\'iquidos y Sistemas Biol\'ogicos, CCT La Plata, CONICET and Departamento de F\'isica, Facultad de Ciencias Exactas, Universidad Nacional de La Plata, C.C. 67, 1900 La Plata, Argentina}
 \affiliation{Abdus Salam International Centre for Theoretical Physics, Associate Scheme, Strada Costiera 11, 34151, Trieste, Italy}
\begin{abstract}
Magnetic skyrmions, topological spin textures observed in chiral magnets, have attracted huge interest due to their applications in the field of spintronics. In this work we study the stability of circular isolated skyrmions in ferromagnetic chiral magnets under the influence of different perturbations and external fields. 
To this end we develop a general systematic procedure based in a harmonic expansion series of the skyrmion boundary which allows the  identifycation of  the breakdown of the skyrmion circular shape on each instability channel independently. We apply our approach to a few representative spin models with actual interest in order to obtain the zero temperature phase diagram, where isolated skyrmions emerge as metaestable states. The results presented in this paper are in agreement with properties of isolated skyrmions observed in recent experiments opening the possibility of extending the analysis to more complex situations.
\end{abstract}

\maketitle

\section{INTRODUCTION}
\label{sec:intro}

In the last years the study of magnetic skyrmions, localized spin textures with nontrivial global topological properties, has been the focus of intense research due to their many interesting properties such as small size (in the scale of nanometers)\cite{NagaosaTokura2013} and, as a consequence of its topological nature, high stability and emergent electrodynamics. 

These properties and their technological potential, gave rise to the development and fast growth of the field of skyrmionics. 
From the initial discovery of skyrmions in the A-phase of the MnSi chiral magnet  \cite{MBJ_09,IA_84,LHP_95} their presence has been reported in a wide variety of systems including metals, semiconductors and insulators and some examples of them are the compounds Mn$_{1-x}$Fe$_x$Ge \cite{SYH_13}, FeGe \cite{LBF_89,UNH_08,YKO_11,WBS_11}, Fe$_{1-x}$Co$_x$Si \cite{BVR_83,GDM_07,GCD_09,OTT_05}, GaV$_4$S$_8$ \cite{KEZ_15}, Cu$_{2}$SeO$_{3}$ \cite{SEKI_12,ADAM_12} among others. 

On the one hand, such magnetic textures with vortex-like structure, were predicted to exist and crystallize in chiral magnets \cite{BY_89,BH_94,BH2_94,RBP_06,ROSS_11,LEO_16} without inversion symmetry. In particular, these objects can be stabilized through a variety of mechanisms as frustrated exchange, dipolar interactions and antisymmetric Dzyaloshiskii-Moriya interaction (DMI)\cite{NagaosaTokura2013} which are found in a large variety of magnetic materials. On the other hand, numerous efforts have been devoted to the manipulations of isolated skyrmions, in particular, controlling their motion would allow for potential technological applications. In this direction, recent research has shown that skyrmions can be manipulated in different ways by means of external fields \cite{OKAMURA_16,WHITE_14} and currents \cite{JIA_16,WOO_16}  even at room temperature. 
These manipulations open the question of how a given perturbation or interaction may turn a skyrmion into an unstable state and eventually change its structure.  
A first study of  skyrmion stability was performed by Bogdanov and Hubert who analyzed the elliptical instability of a skyrmion \cite{BH2_94}. With a similar spirit but in the context of magnetic bubbles,  Thiele\cite{THI_69, THI_70}  studied the stability of a cylindrical magnetic bubble under arbitrary shape deformations. However, a general approach to study skyrmion stability under arbitrary deformations and/or interactions is still lacking and it is the main aim of our work.  

In this paper we develop a method to systematically analyze circular stability under arbitrary deformations of isolated skyrmions in two-dimensional spin system. In particular, this approach allows for the study of skyrmions where the shape instability is driven by microscopic anisotropies or external fields, as well as the study of the manipulation of the skyrmion by means of external electric and magnetic fields.

In order to test our framework, we apply the method to some realistic spin models. First, we consider both the isotropic and anisotropic DMI models\cite{SHI_15} which provide the starting point for the stability analysis of the subsequent models. Then, in order to study more relevant situations, we analyze the effect of a tilted magnetic field and an electric field through the magneto-electric coupling. In these situations, our analysis shows that for large fields and moderate perturbations it is possible to destabilize a circular skyrmion by tuning the magnetic field. In addition, our results show that depending on the strength of the interaction the instability can lead to cardioid, elliptic or triangular deformations.

The paper is organized as follows: in Sec. \ref{sec:method} we present the general framework for skyrmion stability and a detailed derivation of the method. 
In Sec. \ref{sec:applications} we apply the method to a series of spin models: isotropic and anisotropic DMI (Sec. \ref{subsec:DMI}), tilted magnetic field (Sec. \ref{subsec:tilted}) and magneto-electric coupling (Sec. \ref{subsec:magneto}). In section \ref{sec:conc} we summarize and discuss the results of the work.

\section{SHAPE INSTABILITY OF CIRCULAR SKYRMIONS}
\label{sec:method}

The low energy description of a general bidimensional magnetic system can be characterized a local magnetization represented by a three-component unimodular vector field ($|\mathbf{m}({\bf r})|=1$) at each site. This vector field represents a mapping from the base space $\mathds{R}^2$ to the target space $S^2$ of the magnetization field. For fields that take a constant values at infinity the base space can be compactified to $S^2$, and the maps can be classified by the second homotopy group $\Pi_2(S_2)\approx \mathds{Z}$ through a topological invariant defined as \cite{NagaosaTokura2013}:

\be
\frac{1}{4\pi}\int d^2r\, \mathbf{m}({\bf r})\cdot(\partial_{x}\mathbf{m}({\bf r}))\times(\partial_{y}\mathbf{m}({\bf r}))=Q\in \mathds{Z}.
\ee

In this way, a non-trivial magnetic field configuration is characterized by a conserved topological charge, called skyrmion charge. Within this classification, a magnetic skyrmion (antiskyrmion) is a particular configuration where $Q>0$ ($Q<0$).

\begin{figure}[htb]
\subfigure{
\includegraphics[width=4cm]{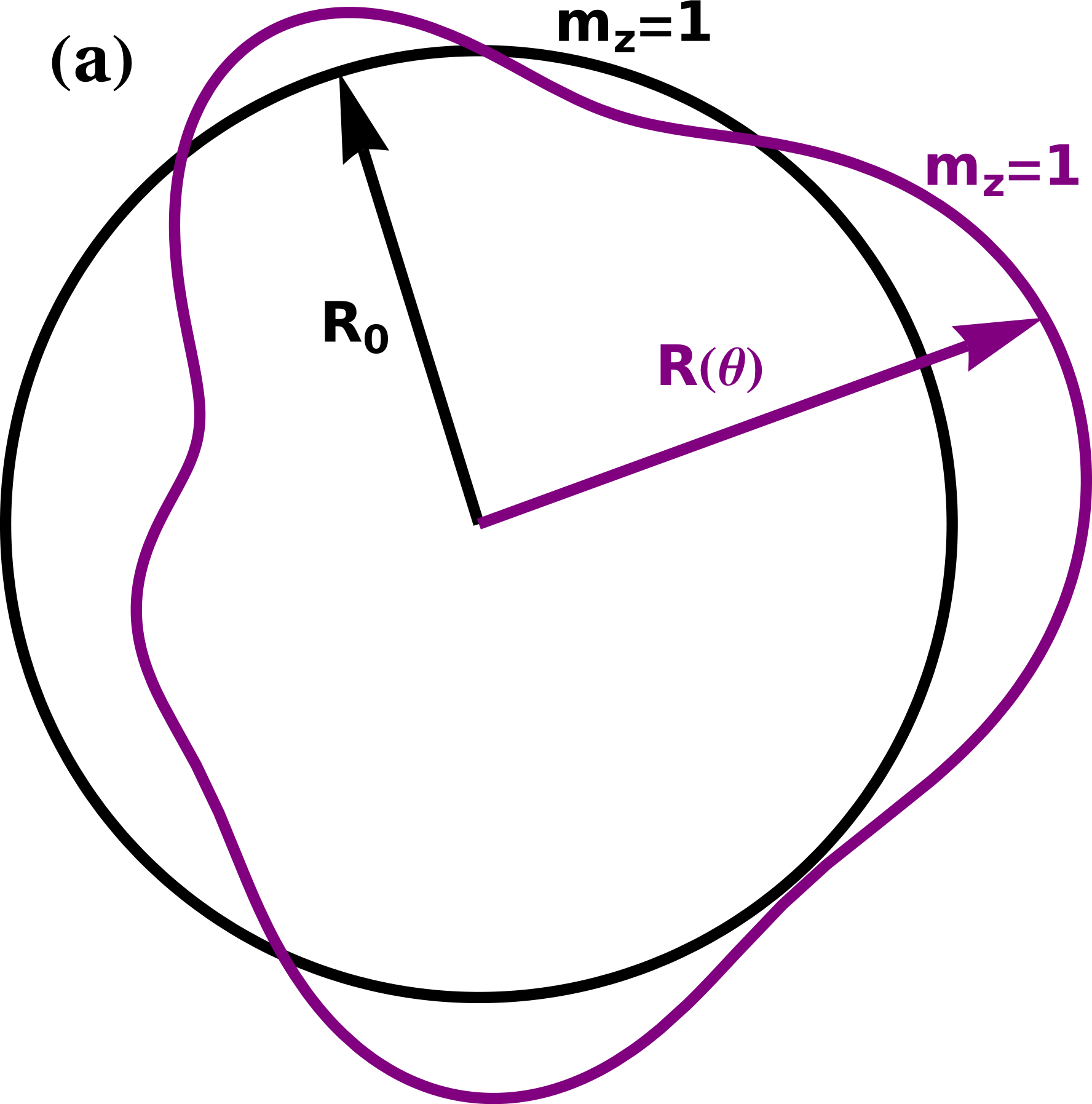}
\label{fig:SkyrmionSchemea}
}
\subfigure{
\includegraphics[width=8cm]{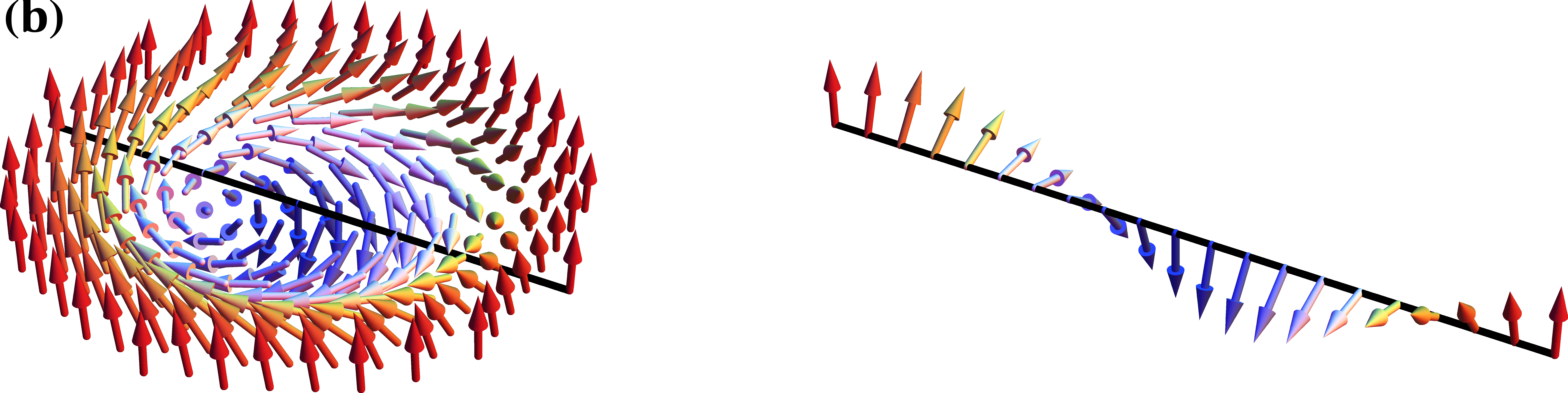}
\label{fig:SkyrmionSchemeb}
}
\subfigure{
\includegraphics[width=8cm]{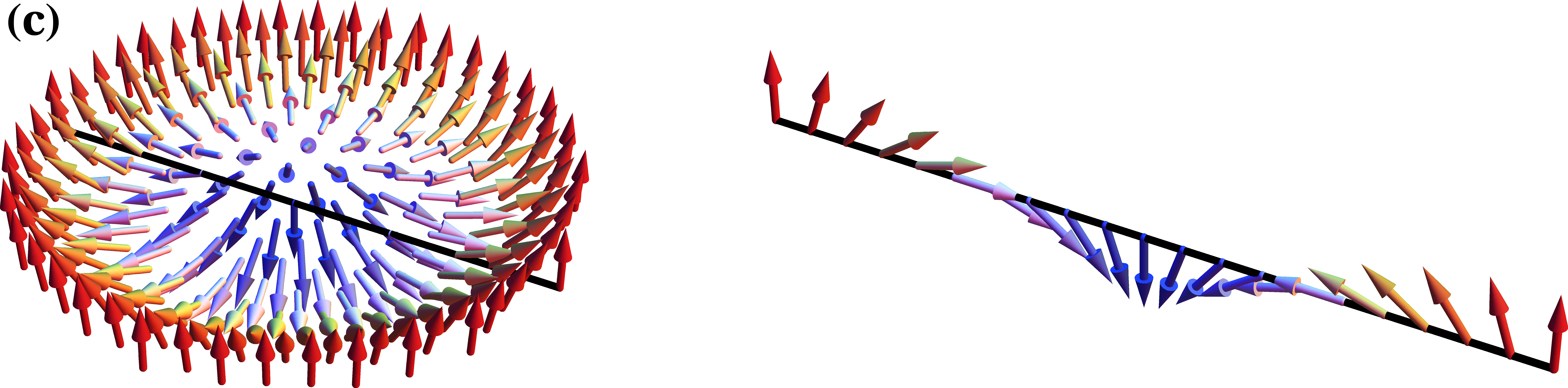}
\label{fig:SkyrmionSchemec}
}
\caption{Skyrmion shape parametrization and skyrmion types. a) The black and purple lines represent contours of constant $m_{z}$ in the case of an axisymmetric skyrmion and a skyrmion of arbitrary shape respectively. A representation of b) a Bloch skyrmion and c) a N\'eel skyrmion.}
\label{fig:SkyrmionScheme}
\end{figure}

In order to study the shape instability of skyrmions, it is convenient to use polar coordinates so that a point ${\bf r}$ in the plane is denoted
by $r$ and $\phi$. The magnetization field of a skyrmion  $\mathbf{m}({\bf r})$ is characterized by two (constant) parameters representing the radius ($R$, defined as the distance from the center to the contour where $m_{z}=1$) of the skyrmion and the in-plane orientation ($\chi$, usually termed as ``helicity'' \cite{NagaosaTokura2013}) of the magnetization.  The swirling structure of a skyrmion centered at $r=0$ can be  described by the finite size skyrmion Ansatz \cite{BY_89}:
\bea
\mv(\rv)&=&(\sin\Theta\cos\Phi,\sin\Theta\cos\Phi,\cos\Theta),
\label{ec:compmag}
\eea
where $\Theta=\Theta(r)=f(r/R)$, $\Phi=\phi+\chi$ and $f(r/R)$ is
\[ f(r/R)=\begin{cases}
      \pi(1-\frac{r}{R}) & r\leq R \\
      0 & r> R.
   \end{cases}
\]
For an arbitrary shape skyrmion (see Fig. \ref{fig:SkyrmionSchemea}), both $R$ and $\chi$ depend on the polar angle $\phi$. So, the radius can be written as,
\be
R(\phi)=R_0+\sum_{n=1}^{\infty}R_{n}\cos(n \phi + \beta_{n}),
\label{ec:curvafourier}
\ee
where $R_{0}$ represents the equilibrium radius of a skyrmion and the second term represents the small deformations written as an expansion in harmonics with coefficients $R_n\ll R_0$. 

The angle between the in-plane component of the magnetization and the tangent to the curve defined by $m_{z}=\text{const.}$ remains approximately constant along the contour \cite{TCH_12}. In order to preserve this condition in a deformed skyrmion we must replace the parameter $\chi$ with a function $\chi(\phi)$. For small perturbations of the skyrmion shape, $\chi(\phi)$ is given up to second order in perturbations (i. e., up to second order in $R_n$, see Appendix \ref{app:chi}) by:

\be
\chi(\phi)=\chi_{0}-\frac{\delta R'(\phi)}{R_{0}}+\frac{\delta R(\phi)\delta R'(\phi)}{R_{0}^{2}},
\label{ec:condtchernysov}
\ee
where $\chi_{0}$ is the orientation of the equilibrium configuration. The case $\chi_{0}=\pm\pi/2$ corresponds to Bloch skyrmions (see Fig. \ref{fig:SkyrmionSchemeb}), where the $xy$ components are parallel to the contour $m_{z}=\text{const.}$ with a clockwise or counterclockwise swirling of the spins; while $\chi_0=0$, corresponds to the N\'eel skyrmions (see Fig. \ref{fig:SkyrmionSchemec}) where the $xy$ spin components are perpendicular to the $m_{z}=\text{const}$.

The method to search for skyrmion instabilities, can be summarized in the following steps:
\begin{enumerate}
 \item Assuming that each spin varies slowly we perform a gradient expansion of the microscopic spin Hamiltonian. 

\item We calculate the energy of a single skyrmion as:

\be
E_{Sk}=\int_{0}^{2\pi}d\phi\int_{0}^{\infty}r\,\mathcal{H}(r,\phi)\,dr,
\ee
where $\mathcal{H}(r,\phi)$ is the energy density evaluated in the previous Ansatz, and subtract from it the energy of the field polarized state
\be
E_{FP}=\int_{0}^{2\pi}d\phi\int_{0}^{\infty}r\,B\,dr,
\ee
which gives
\be
E=E_{Sk}-E_{FP}=\int_{0}^{2\pi}d\phi\int_{0}^{R(\phi)}r\,\mathcal{H}(r,\phi)\,dr.
\label{eq:DofEnergy}
\ee

\item We expand the energy up to second order in $R_{n}$ around the circular configuration (with radius $R_{0}$ and helicity $\chi_{0}$) and we obtain an expression that takes the general form:

\bea
E&=&E_0+\sum_{n=1}^{\infty}L_n\,R_n+\sum_{n,m=1}^{\infty}R_n\,M_{nm}R_m,
\label{eq:desarrolloGeneral}
\eea
where $E_{0}$ represents the energy of the axisymmetric skyrmion, $L_{n}$ and $M_{nm}$ are coefficients expressed in terms of the microscopic parameters (spin coupling constants) and external parameters (couplings to external fields) of the model. 

\item The stability analysis involves the study of the sign of the eigenvalues of the matrix $M_{nm}$. These eigenvalues are functions of the microscopic parameters and the external fields. The stability condition requires positive eigenvalues of $M_{nm}$ wich in turn determines the phase diagram where a skyrmion can exists as a metastable state. Finally we determine the shape of the skyrmion, in the stability region, by minimizing the energy with respect to $R_{n}$, 

\be
R_n=-\frac{1}{2}\sum_{m=1}^{\infty}M^{-1}_{n m}L_{m}.
\label{eq:eclineal}
\ee

\end{enumerate}

A proper treatment of the stability analysis requires to consider the full series of harmonics in the expansion Eq. (\ref{ec:curvafourier}) which leads to the diagonalization problem of the matrix $M$ in Eq. (\ref{eq:desarrolloGeneral}) that may seem at first sight rather involved. However, we should remark some important properties of the matrix $M$ which simplifies the analysis. In the first place the matrix $M$, within the continuum limit, has to be considered as a finite dimensional matrix. To see this we should note that higher harmonics represent short length fluctuations of the spin field, then, the validity of the expansion Eq. (\ref{ec:curvafourier}) is restricted by the plausibility of the continuum description. Thus, for a sufficiently high value (say $N$) of the index $n$ the continuous description breaks down. This introduces a natural cutoff that limits the sum up to the first $N$ terms. In the second place, from our approach we find that the general structure of the matrix $M$ corresponds to a  symmetric band matrix. Although this problem could be studied numerically, in order to keep our analysis in its simplest form and to provide analytical results, we consider the first $n\leq4$ deformations in the expansion Eq. (\ref{ec:curvafourier}). Then the problem is reduced to the study of a square matrix of lower dimension. As we show for the models considered here, those few harmonics are sufficient for a phenomenological description, which makes our method a powerful tool even in its simplest form.

We are going to apply the method to several models of interest.
In the first place we consider the isotropic and anisotropic DMI model \cite{SHI_15}. Then we study the effect of the magnetic field tilting \cite{LIN_15}. Finally we study the effect of the magneto-electric coupling of the spins to an external electric field \cite{MOCHI_13, MOCHI_15, MOCHI_16}.

\section{Models}
\label{sec:applications}
\subsection{Dzyaloshinskii-Moriya interaction}
\label{subsec:DMI}

The first example where we are going to apply the method corresponds to the nearest-neighbor ferromagnetic Heisenberg model  including anisotropic DMI\cite{SHI_15}, so the microscopic Hamiltonian, for the square lattice, is written in the following form:

\bea
H&=&-\sum_{\rv,\alpha}\left\lbrace J\Sp_{\rv}\cdot\Sp_{\rv+\dv_\alpha}+D_\alpha \delta_{\alpha}\cdot(\Sp_\rv\times\Sp_{\rv+\dv_\alpha})\right\rbrace\nonumber\\
&&-B\sum_{\rv}S^{z}_\rv,
\eea
where $\alpha=x,y$,  $\dv_{\alpha}$ is the unit vcector along the $\alpha$ direction ($\delta_{x}=\mathbf{\hat{x}}$, $\delta_{y}=\mathbf{\hat{y}}$) and $\mathbf{S}_{\mathbf{r}}$ are classical spins on site $\mathbf{r}$. In the previous equation the anisotropic DMI term is introduced by adopting different coefficient $D_x$ and $D_y$ for the $x$ and $y$ directions, respectively. The magnetic field is in the $z$ axis and perpendicular to the film.

In the continuum limit, for a two dimensional system, the energy density is:
\bea
\nonumber
\mathcal{H}&=&\sum_{\alpha=x,y}\left\lbrace\frac{J}{2}(\partial_{\alpha}\mv)^2+D_{\alpha}(\delta_\alpha (\mv\times\partial_{\alpha}\mv)\right\rbrace-\\
&-&B\,m_{z}.
\label{eq:DofEnergyModel1}
\eea
Using the Eqs.(\ref{ec:compmag})-(\ref{ec:condtchernysov}) and (\ref{eq:DofEnergyModel1}) in Eq. (\ref{eq:DofEnergy}), the different energy terms are given by:
\bea
E^{\nu}&=&E_{0}^{\nu}+\sum_{n=1}^{\infty}L_{n}^{\nu}R_{n}+\sum_{n,m=1}^{\infty}(M_{n m}^{\nu}R_{n}R_{m}),
\label{eq:EnergyGeneralCase}
\eea
where $\nu$ labels the isotropic ($I$) and the anistropic cases ($A$) that will be analyzed in this section. The coefficients $E_0^{\nu}, L_n^{\nu}, M_{n m}^{\nu}$, depend on $J$, $D^{\pm}=D_y\pm D_x, B$  and $R_0,\chi_0$ with explicit expressions given in Appendix \ref{app:constants}. The vector $L_{n}^{\nu}$ and the matrix $M_{n m}^{\nu}$ depend upon $\beta_n$ as well. The $R_{0}$ and $\chi_{0}$ parameters, related to the non-perturbed skyrmion case, are obtained by direct minimization of the total energy  setting $R_{n}=0, \forall n$ in Eq. (\ref{eq:EnergyGeneralCase}) (in the general case we set $R_{n}=0$ in Eq. (\ref{eq:desarrolloGeneral}) and calculate $R_{0}$ and $\chi_{0}$). Finally, we determine the stability of a given configuration using these values in Eq. (\ref{eq:EnergyGeneralCase}).

We are going to consider two cases of the previous model:\\
\begin{itemize}
 \item \emph{Isotropic case, $D_{x}=D_{y}$} ($D^{-}=0$): it is known that axisymmetric (circular) skyrmions are present in this system. 
 \item \emph{Anisotropic case, $D_{x}\neq D_{y}$} ($D^{-}\neq0$): in this case the skyrmions are elliptical and crystallize in a distorted triangular lattice along the strain direction as showed in Shibata et. al \cite{SHI_15}.
\end{itemize}
\subsubsection{Isotropic case}
\label{subsubsec:iso}

We start with the \emph{isotropic} model setting $D_{x}=D_{y}=D$ in the Eq. (\ref{eq:EnergyGeneralCase}). This case is very important because we are going to analyze the effect of anisotropy, tilting of a magnetic field and electric field as a perturbation around the circular skyrmion.
In the isotropic case we have $D^-=0$ and the energy has a very simple expression given by:

\bea
E^{I}&=&E_0^{I}+\sum_{n,m=1}^{\infty}\tilde{M}_{n m}^{I}R_n R_m.
\label{eq:EnergyCase1}
\eea
The matrix $\tilde{M}_{n m}^{I}$ doesn't depend on $\beta_{n}$ and the equilibrium parameters $R_{0}$ and $\chi_0$ are (for $B>0$):
\bea
R_{0}=\frac{\Lambda_{4}|D|}{2\Lambda_{3}B}&,&\chi_{0}=-sign(D)\frac{\pi}{2},
\label{eq:pareq}
\eea
where $sign(D)$ returns the sign of $D$. 

\begin{figure}[!]
\subfigure{
\includegraphics[width=4cm]{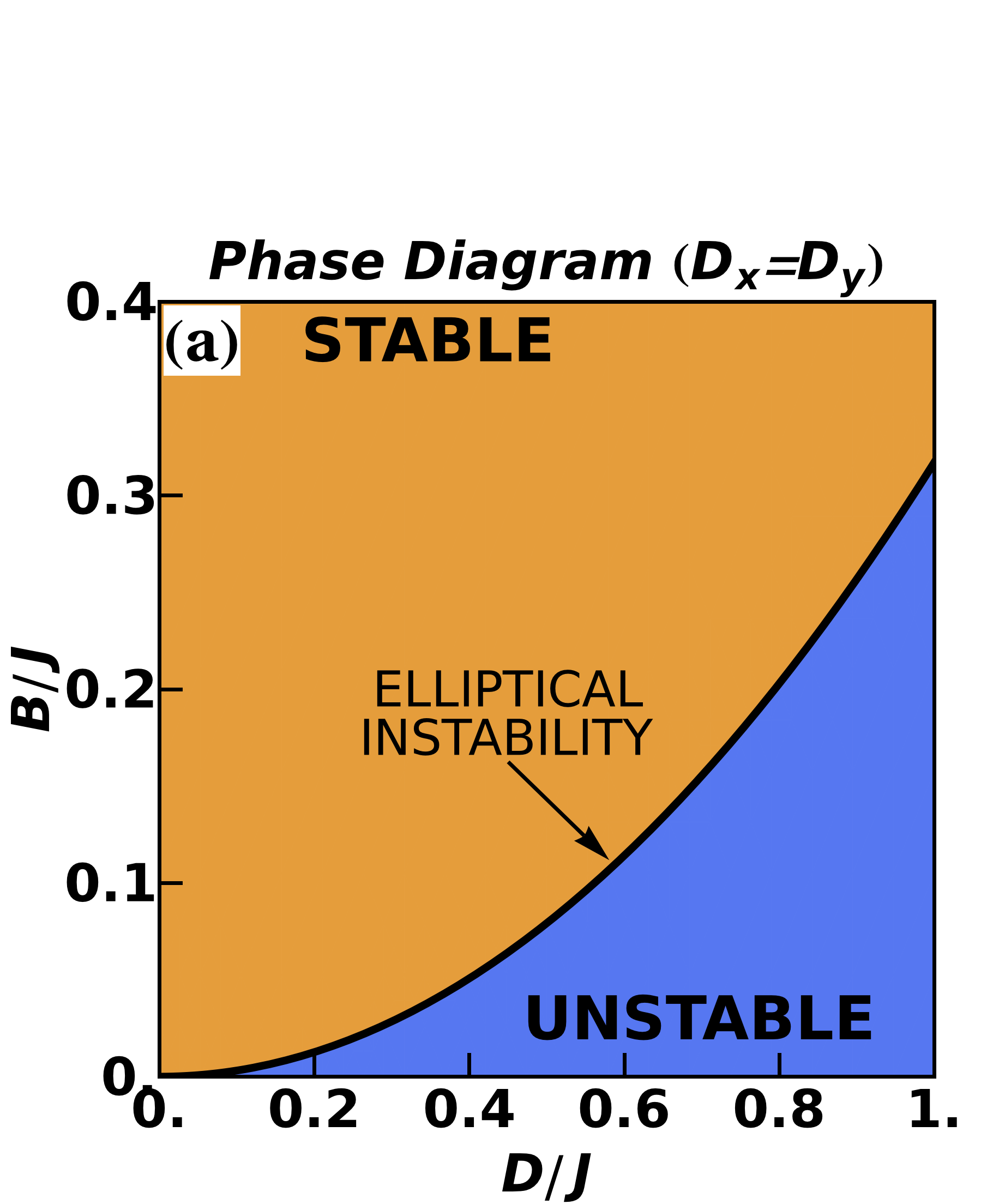}
\label{subfig:isot}
}
\subfigure{
\includegraphics[width=4cm]{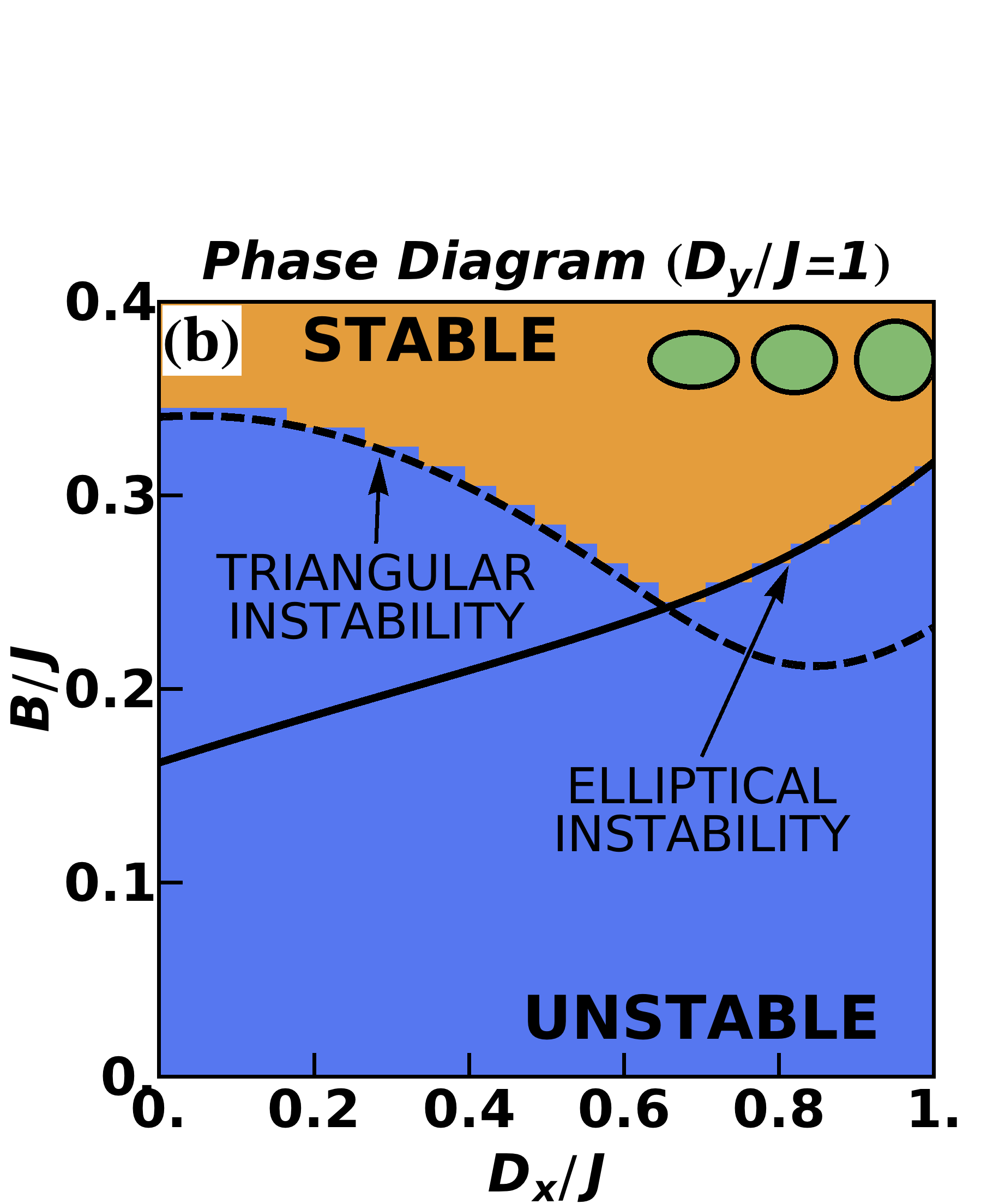}
\label{subfig:anisot}
}
\subfigure{
\includegraphics[width=4cm]{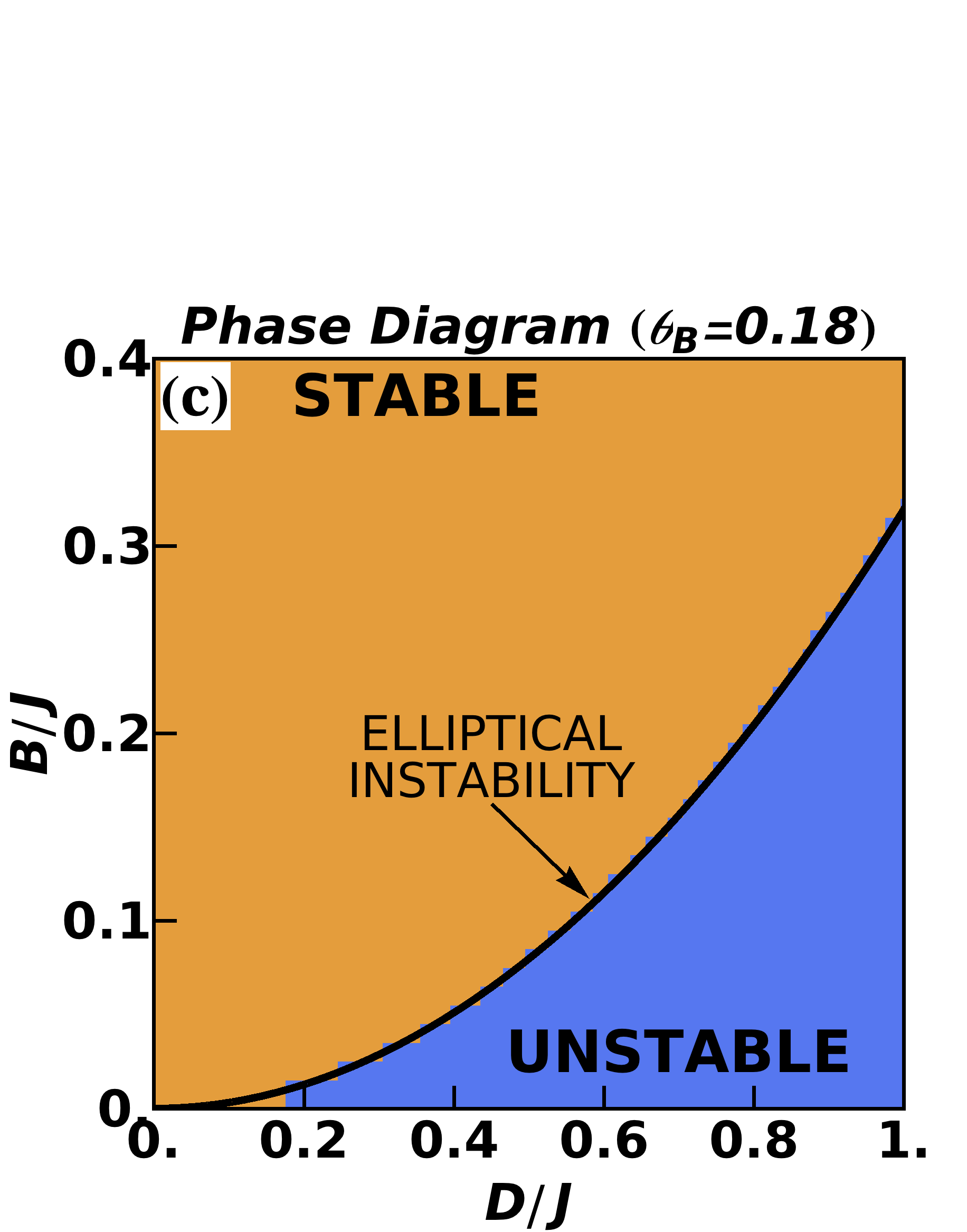}
\label{subfig:tilted}
}
\subfigure{
\includegraphics[width=4cm]{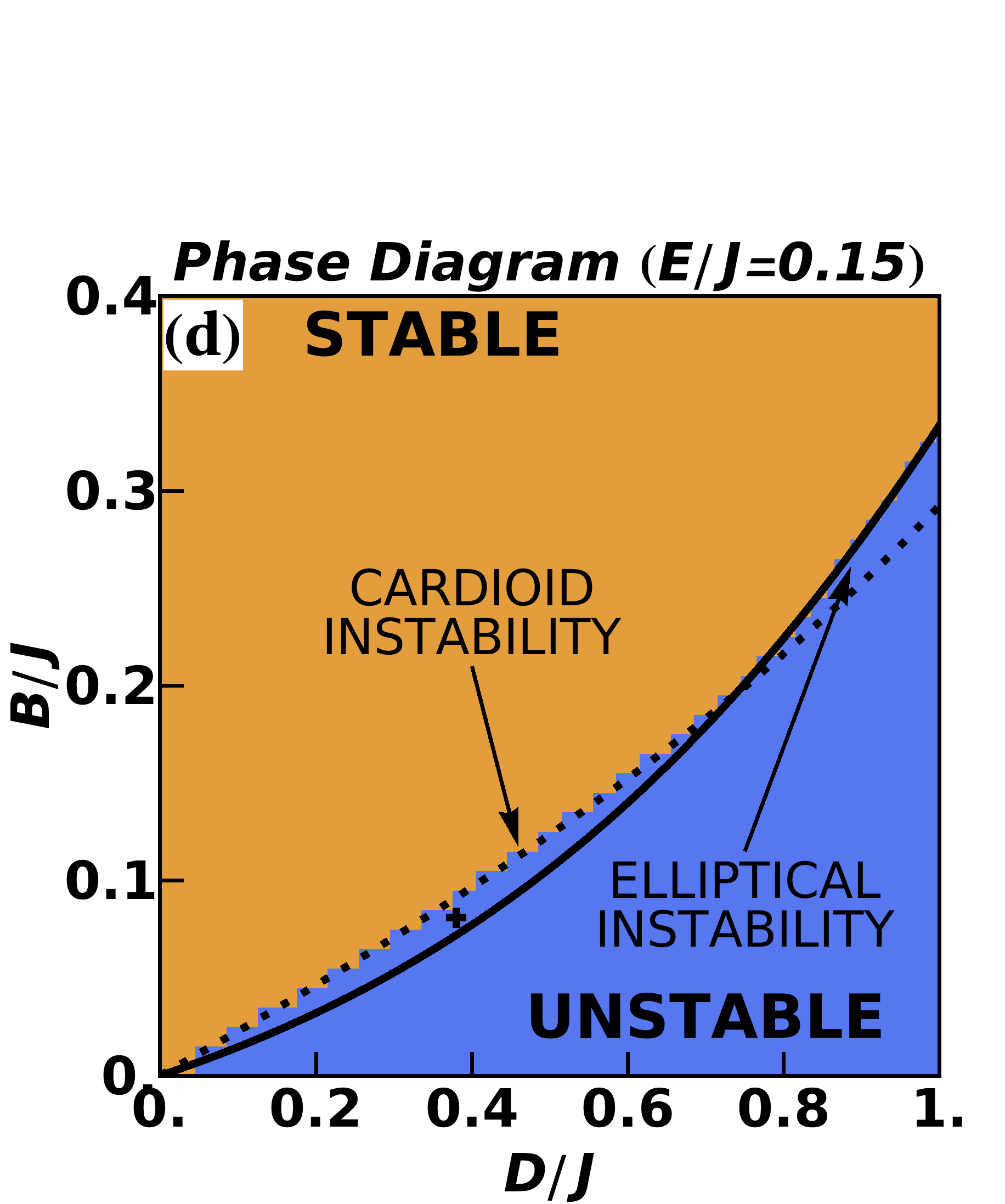}
\label{subfig:magnetoelectricperp}
}
\caption{Stability phase diagrams of isolated skyrmions for the models studied here. In the orange region skyrmions are stable and in the blue one they are unstable.(a) Isotropic DMI model in a transverse magnetic field: The solid curve is the critical field $B^{I}_{2}$. (b) Anisotropic DMI model (for $D_{y}/J=1$) in a transverse magnetic field: The solid curve represents $B^{A}_{2}$ and the dashed one $B^{A}_{3}$. The colored background (orange and blue) corresponds to the numerical results for 100 modes. In green we represent the distortion of a skyrmion from a circular (on the right) to an elliptical skyrmion. (c) Isotropic DMI model in a tilted magnetic field: The solid line is the critical field $B_{2}^{ZT}(\theta_{B})$ for $\theta_{B}=0.18$. (d) Isotropic DMI model in a transverse electric ($E/J=0.15$) and magnetic fields. The lines are the critical magnetic fields $B_{1}^{MEP}$ (dotted) and $B_{2}^{MEP}$ (solid).}
\label{fig:stabcirc}
\end{figure}

The matrix $\tilde{M}_{n m}^{I}=\lambda^{I}_{n}\delta_{n m}$ is diagonal with eigenvalues
\be
\lambda^{I}_{n}=\pi\left[\frac{J}{2}(\Lambda_{1}n^{2}+\Lambda_{2}n^{4})\left(\frac{2\Lambda_{3}B}{\Lambda_{4}D}\right)^{2}+\left(n^{2}-1\right)B\Lambda_{3}\right],
\label{ec:circauto}
\ee
the explicit values for the constants $\Lambda_i$ are given in the Appendix \ref{app:constants}.  
The stability under a given deformation requires $\lambda^{I}_{n}>0$.
This condition introduces a set of critical fields of the form $B^{I}_{n}= F_{n}D^2$,
\be
F_{n}=\frac{2\Lambda_{3}(1-n^{2})}{J(\Lambda_{1}n^{2}+\Lambda_{2}n^{4})}(\frac{\Lambda_{4}}{2\Lambda_{3}})^2,
\ee
thus, for $B<B^{I}_{n}$ the $n$-th mode becomes unstable. An important consequence is that for the $n=1$ mode $F_{1}=0$ and then the first mode is stable for all values of $B>0$. It is easy to see that for $n\ge2$, $F_{n}>F_{n+1}$. This defines the region of stability of the skyrmions in the phase space as shown in Fig. \ref{subfig:isot}. So, the first unstable mode corresponds to $n=2$. The corresponding critical field is given by

\be
B^{I}_{2}=\frac{-6\Lambda_{3}}{J(4\Lambda_{1}+16\Lambda_{2})}(\frac{\Lambda_{4}}{2\Lambda_{3}})^2 D^{2}.
\label{ec:campocrit}
\ee

It is important to note that in  Eq. (\ref{eq:EnergyCase1}) there are no linear terms in the energy ($\tilde{L}_{n}^{I}=0$). In consequence $R_{n}=0,\forall n$ and the stable configuration corresponds to the circular skyrmion (Fig. \ref{subfig:fa}). Then the condition $|\delta R(\phi)|\ll R_0$ is trivially satisfied.

In view of our result the instability of a skyrmion is driven by an elliptical deformation. This suggests that the skyrmion becomes a helical domain as explained by the theory developed in \cite{BH2_94}. This ``strip-out'' process was observed in chiral skyrmions \cite{BH2_94,  LEO_16} as well as in magnetic bubble domains \cite{THI_70,HUB_SCH_98,BRAN_09}. The critical field that we obtain for the stability of a skyrmion is $B^{I}_{2}(D)\approx0.317 D^{2}$ and this value is consistent with the critical field ($B_{H}(D)\approx0.20 D^{2}$) for the transition from the skyrmion crystal (SkX) to the helix state (H) reported by Han et. al \cite{HZY_10}.

\subsubsection{Anisotropic case}
\label{subsubsec:aniso}

Now we turn our attention to the \emph{anisotropic} case ($D_x\neq D_y$) of the Eq. (\ref{eq:EnergyGeneralCase}). There are three important issues of the previous equation to be discussed. In the first place the anisotropy introduces couplings between the $n$ and $n+2$ deformations, in second place the energy depends on the values of $\beta_{n}$ (see Eq. (\ref{ec:curvafourier})), and in third place, there is a linear term in $R_2$ (in this case $L_{n}^{A}=L_{2}^{A}\delta_{n 2}$, see Appendix \ref{app:constants}). The first point represents a change in the quadratic part to a non-diagonal quadratic form but, in this case, the study of stability is a simple matter of diagonalization. The second point is a consequence of the anisotropy of the model, so different orientations of the skyrmion will have different energies. However the most important point is the third, which strongly suggests the presence of a finite value of $R_2$, that is to say an elliptical deformation of the skyrmions.

The value of $\beta_{n}$ that minimizes the energy (assuming $D^{-}>0$ and $D^{+}>0$) satisfies the equations:
\bea
\chi_{0}-2\beta_{1}&=&\pi/2,\nonumber\\
\beta_{2}-\chi_{0}&=&\pi/2,\nonumber\\
\beta_{n}-\beta_{n+2}+\chi_{0}&=&3\pi/2, \forall n>2,
\label{eq:betasaniso}
\eea
where the equilibrium parameters $R_{0}=\frac{\Lambda_{4}D^{+}}{4\Lambda_{3}B}$, $\chi_{0}=-\frac{\pi}{2}$. From previous equations we see $\beta_{1}=3\pi/2$, $\beta_{2}=0$ and $\beta_{n+2}=\beta_{n} + 2 k_{n}\pi, k_{n}\in \mathbb{Z}$, so we have the general form for $\beta_{n}=\frac{3\pi}{2}(\frac{1+(-1)^{n+1}}{2})$.

Hence we see that the energy of the skyrmion as a function of $R$'s is given by
\be
E^{A}=E_{0}^{A}+\tilde{L}_2^{A} R_2+\sum_{n,m=1}^{\infty}R_{n}\tilde{M}_{nm}^{A}R_{m},
\ee
where $\tilde{L}_2^{A}$ and $\tilde{M}_{nm}^{A}$ are the vector $L_n^{A}$ and matrix $M_{nm}^{A}$ with $\beta_n$ evaluated in the values determined through Eq. (\ref{eq:betasaniso}). As we previously mentioned, the instability of the circular phase will depend on identifying the negative eigenvalues of the matrix $\tilde{M}^{A}$. 
We are going to consider just the first four modes in the expansion Eq. (\ref{ec:curvafourier}) to compute the eigenvalues of the matrix $M^{A}$ (of dimension $4 \times 4$). Within this simplification the critical fields and deformations can be calculated analytically. The expressions obtained are rather involved to include them in the text (a situation that is repeated in the rest of the models studied). However, they are very important to determine the general behavior of the critical fields and deformations. It also allows us to test the degree of convergence of the method. At the end of this section we will show that these analytical results are consistent with the numerical ones for an expansion containing a large number of modes.\\
 
The sign of the eigenvalues depends on the external magnetic field ($B$) and the DM interaction ($D_{x}$ and $D_{y}$). Stability requires $\lambda^{A}_{n}>0$ which defines the stability region in $(D_{x},D_{y},B)$ phase diagram. Then, for fixed $D^{\pm}$ the configuration will be stable when the external magnetic field $B$ is bigger than the critical fields $B^{A}_{2}$ and $B^{A}_{3}$. In the isotropic limit ($D_{x}\to D_{y}$) these critical fields coincide with the critical fields $B^{I}_{2}$ and $B^{I}_{3}$ for the isotropic problem. Thus we refer to $B^{A}_{2}$ and $B^{A}_{3}$ as the critical fields for the elliptical and triangular instabilities respectively.

In Fig. \ref{subfig:anisot} we present the phase diagram for $D_{y}/J=1$. In the region $D_{x}/J \gtrsim 0.66$ the critical field $B^{A}_{2}$ (black solid curve) is bigger than $B^{A}_{3}$ (black dashed curve), while $B^{A}_{3} > B^{A}_{2}$ in the region $D_{x}/J \lesssim 0.66$. For small anisotropies the instability of the skyrmion takes place through an elliptical deformation as occurs in the isotropic case. In the opposite extreme, when $D_{x}=0$, the system behaves as a set of spin chains coupled to each other only by ferromagnetic exchange. In each chain the spins interacts with their neighbors via ferromagnetic exchange and DMI in the $\hat{y}$ direction. From our phase diagram Fig. \ref{subfig:anisot} we see that the skyrmion becomes unstable through a triangular deformation.

The shape of the skyrmion is determined by the values of $R_{1}$, $R_{2}$, $R_{3}$ and $R_{4}$ that minimizes the energy in the stability region.  
In the isotropic case $D^{-}=0$ the skyrmion is circular as we saw in the previous discussion (Sec. \ref{subsubsec:iso}). For $D^{-}\neq0$ the coefficients $R_{1}$ and $R_{3}$ are identically zero, while $R_{2}\propto D^{-}$ and $R_{4}\propto (D^{-})^{2}$. So in the small anisotropy regime the skyrmion is approximately elliptical as shown in Fig. \ref{subfig:fb}. To illustrate how the skyrmion shape depends on $D_{x}$ and $D_{y}$ we represent, in Fig. \ref{subfig:anisot}, the skyrmion shape for $D_{y}/J=1$ and $0 \leq D_{x}/J\leq 1$. As the value of $D_{x}$ decreases the skyrmion elongates along $\hat{x}$ direction essentially through the elliptical deformation. We can compare this result with the results found by Shibata and collaborators  \cite{SHI_15}. In this reference the authors studied FeGe films under uni-axial strain. They have shown that the SkX structure, and also each skyrmion, are deformed along the strain direction. To explain this phenomena they propose an anisotropic DMI induced by thermal strain. Our results on the anisotropic DMI demonstrate how this deformation, and the shape of the skyrmion, depends on the microscopic parameters. 

\begin{figure}[!]
\subfigure{
\includegraphics[width=8cm]{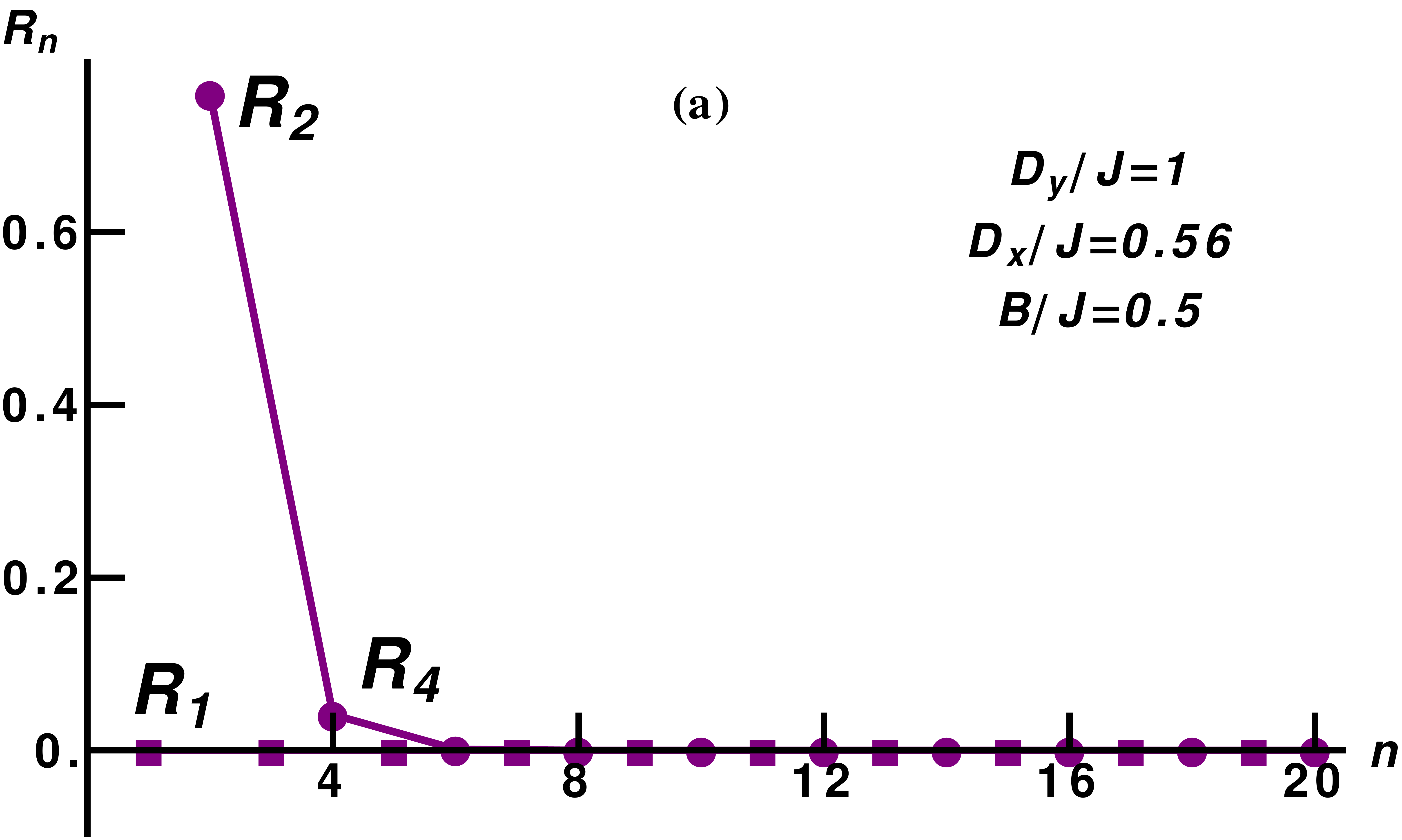}
\label{subfig:ranisot}
}
\subfigure{
\includegraphics[width=8cm]{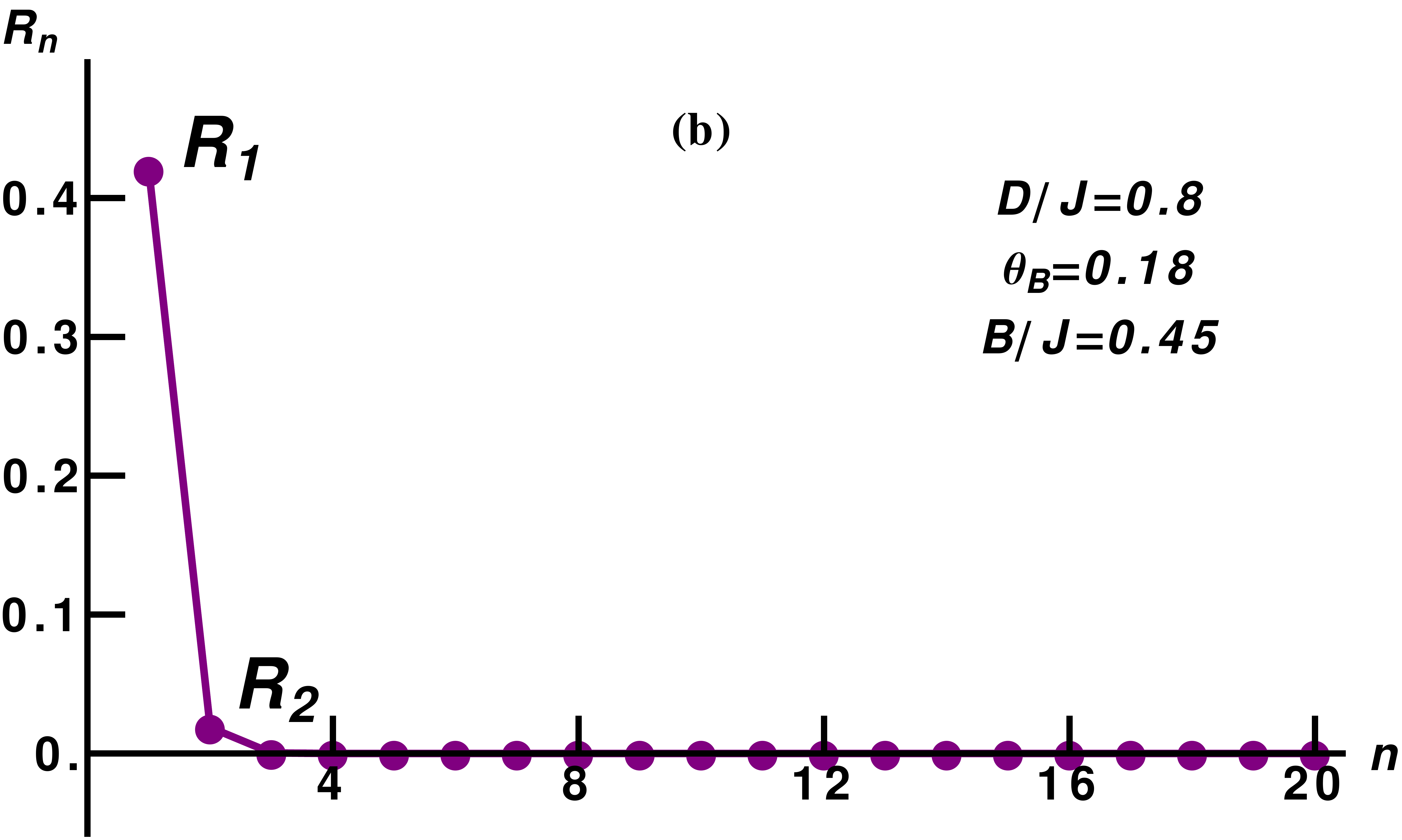}
\label{subfig:rtilted}
}
\subfigure{
\includegraphics[width=8cm]{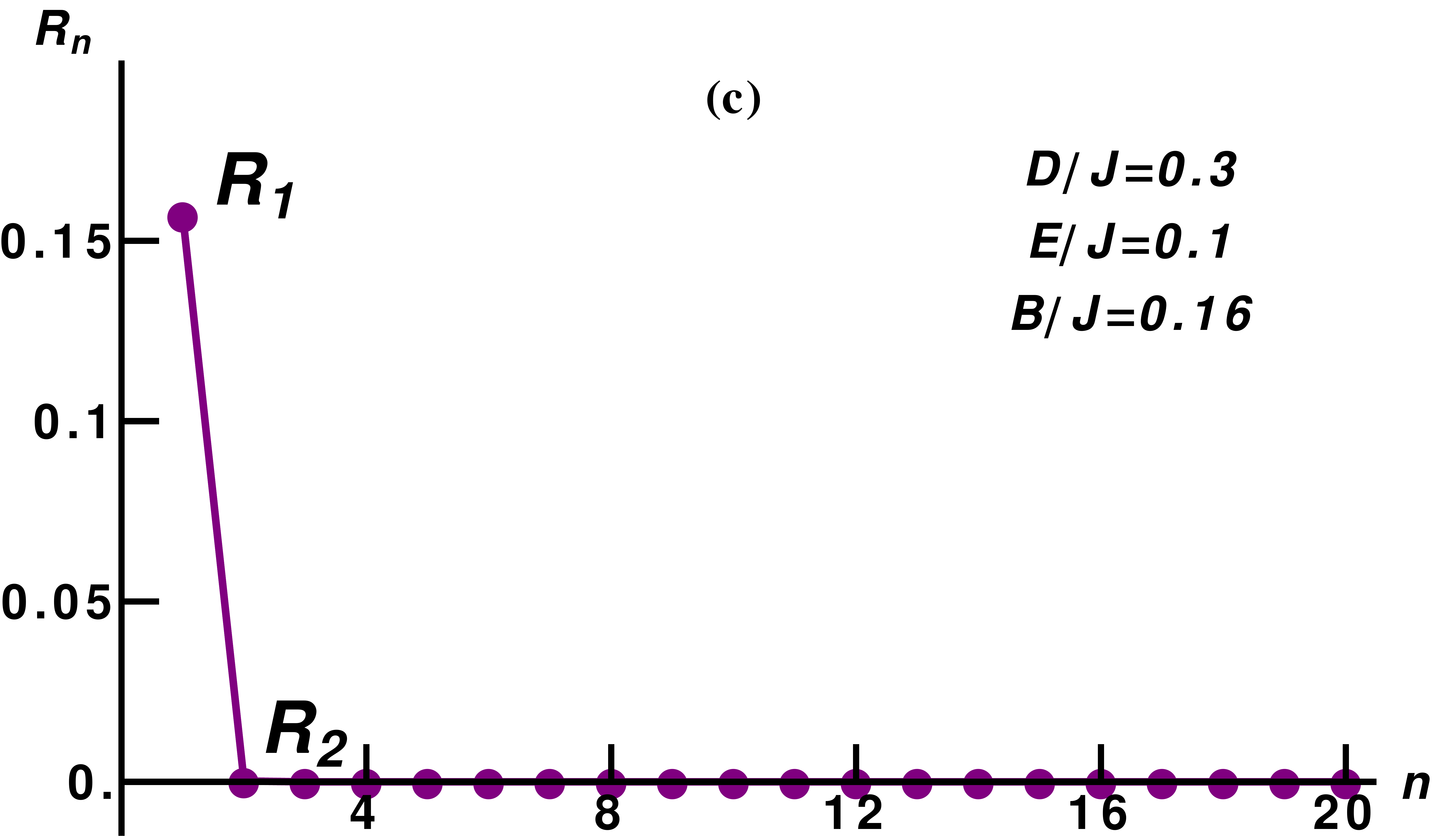}
\label{subfig:rmeperp}
}
\caption{(a) The coefficients $R_{n}$ for the anistoropic DMI model ($D_{y}/J=1$, $D_{x}/J=0.56$ and $B/J=0.5$). The circles (squares) are the values of $R_{n}$ with even (odd) values of $n$. (b) Coefficients $R_{n}$ for the isotropic model in a tilted magnetic field ($D/J=0.8$, $\theta_{B}=0.18$ and $B/J=0.45$). (c) Coeffiecients $R_{n}$ for the isotropic model in an in-plane external electric field ($D/J=0.3$, $E/J=0.1$ and $B/J=0.16$).}
\label{fig:rstabcirc}
\end{figure}

Now we discuss the numerical results.
In Fig. \ref{subfig:anisot} is represented the phase diagram obtained with the numerical approach for a $100$ modes expansion in the Eq. (\ref{ec:curvafourier}). The boundaries between the stable (orange) and unstable region (blue) are delimited by the critical fields $B^{A}_{2}$ and $B^{A}_{3}$, calculated analytically, showing and excellent agreement between both approaches. Another important result of the numerical approach regards to the behaviour of the coefficients $R_{n}$ (see Fig. \ref{subfig:ranisot}). For odd values of $n$, $R_{n}=0$, while for even values, $R_{n}$ is a rapidly decreasing function of $n$. The most relevant deformations are the elliptical ($R_{2}$) and square ($R_{4}$) one, which serves as an auto-consistency checking of both approaches.

\subsection{Tilted magnetic field}
\label{subsec:tilted}

Now we study the effect of tilting the magnetic field in the isotropic system. We parametrize the magnetic field as $\mathbf{B}=B(\sin{(\theta_{B})}\cos{(\phi_{B})},\sin{(\theta_{B})}\sin{(\phi_{B})},\cos{(\theta_{B})})$ with $\phi_B, \theta_B$ azimuthal and polar angles respectively. When the magnetic field is tilted from the perpendicular position the spin align itself with the magnetic field far away from the skyrmion core. So in this case we consider that the spin field for the skyrmion is given by Eq. (\ref{ec:compmag}) with $f(r/R)$ and $\Phi$ changed by: 

\[ f(r/R)=\begin{cases}
      \pi(1-\frac{r}{R}) & r\leq R \\
      \theta_{B} & r> R,
   \end{cases}
\]
\[ \Phi(r)=\begin{cases}
      \phi+\chi & r\leq R \\
      \phi_{B} & r> R.
   \end{cases}
\]
Then, as we mentioned in Sec. \ref{sec:method}, we compute the energy of the skyrmion with respect to the field polarized state (now in the direction $\mathbf{\hat{B}}$). The reader should note that the spin field has a discontinuity at $r=R$. However, assuming a small $\theta_B$ ($\theta_{B}<<1$), the contribution of this discontinuity to the energy is of order $\theta_{B}^{2}$ (see Appendix \ref{app:discon}) and we can neglect it for small inclination angles.

In this limit, all the terms of the energy remain unchanged except the Zeeman term that takes the form:

\be
E^{ZT}=E_{0}^{ZT}+\sum_{n}^{\infty}L_{n}^{ZT}R_{n}+\sum_{n,m=1}^{\infty}R_{n}M_{n m}^{ZT}R_{m},
\ee
the expressions for $E_{0}^{ZT}$, $L_{n}^{ZT}$ and $M_{n m}^{ZT}$ are in the Appendix \ref{app:constants}.

The consequence of a small tilting angle $\theta_{B}$ is the coupling between different deformations amplitudes $R_n$. In addition a linear term proportional to $R_{1}$ emerges allowing a deformation of the skyrmion  
through a finite value of $R_{1}$. For definiteness we consider $0\le \theta_{B}\le \pi/2$ and $0\le\phi_{B}\le \pi$. Under these assumptions the values for $\beta_n$ leading to the minimum of the energy are:

\be
\beta_{n}=n(\chi_{0}-\phi_{B}).
\label{eq:betatilted}
\ee

The determination of the equilibrium parameters ($R_{0}$ and $\chi_{0}$) and the stability analysis of this case goes on the same line as in the isotropic case for perpendicular magnetic field replacing $\Lambda_{3}$ by $\tilde{\Lambda}_{3}$ in Eq. (\ref{eq:pareq}). For definiteness and for later comparison with available simulations \cite{LIN_15} we chose $D<0$ thus $\chi_{0}=\pi/2$.

If we consider just the first two deformations in the expansion Eq. (\ref{ec:curvafourier}) we can diagonalize the Hessian analytically. The positive values  of the igenvalues define the critical field $B_{2}^{ZT}(\theta_{B})$.
This critical field reduces to $B_{2}^{I}$ (Eq. (\ref{ec:campocrit})) when $\theta_{B}\to0$. For this reason we will interpret $B_{2}^{ZT}(\theta_{B})$ as the critical field for the SkX $\to$ H transition as we did in the discussion of the isotropic model in transverse magnetic field. For magnetic fields below this critical field the skyrmion is unstable. The phase diagram (Fig. \ref{subfig:tilted}) is almost identical to the perpendicular magnetic field case. In our analysis we find that $B_{2}^{ZT}(\theta_{B})$ (for small $\theta_{B}$) is an increasing function of $\theta_{B}$. Thus the effect of the tilting is just a slight reduction of the skyrmion stability region. This result can be compared with the phase diagram presented in Ref.\cite{LIN_15}. In this paper a phase diagram for different inclination angles is presented. The authors find that the critical fields for the transition SkX $\to$ H, increases when the magnetic field is tilted. This behaviour is in qualitative agreement with our findings. 

\begin{figure}[!]
\subfigure{
\includegraphics[width=4cm]{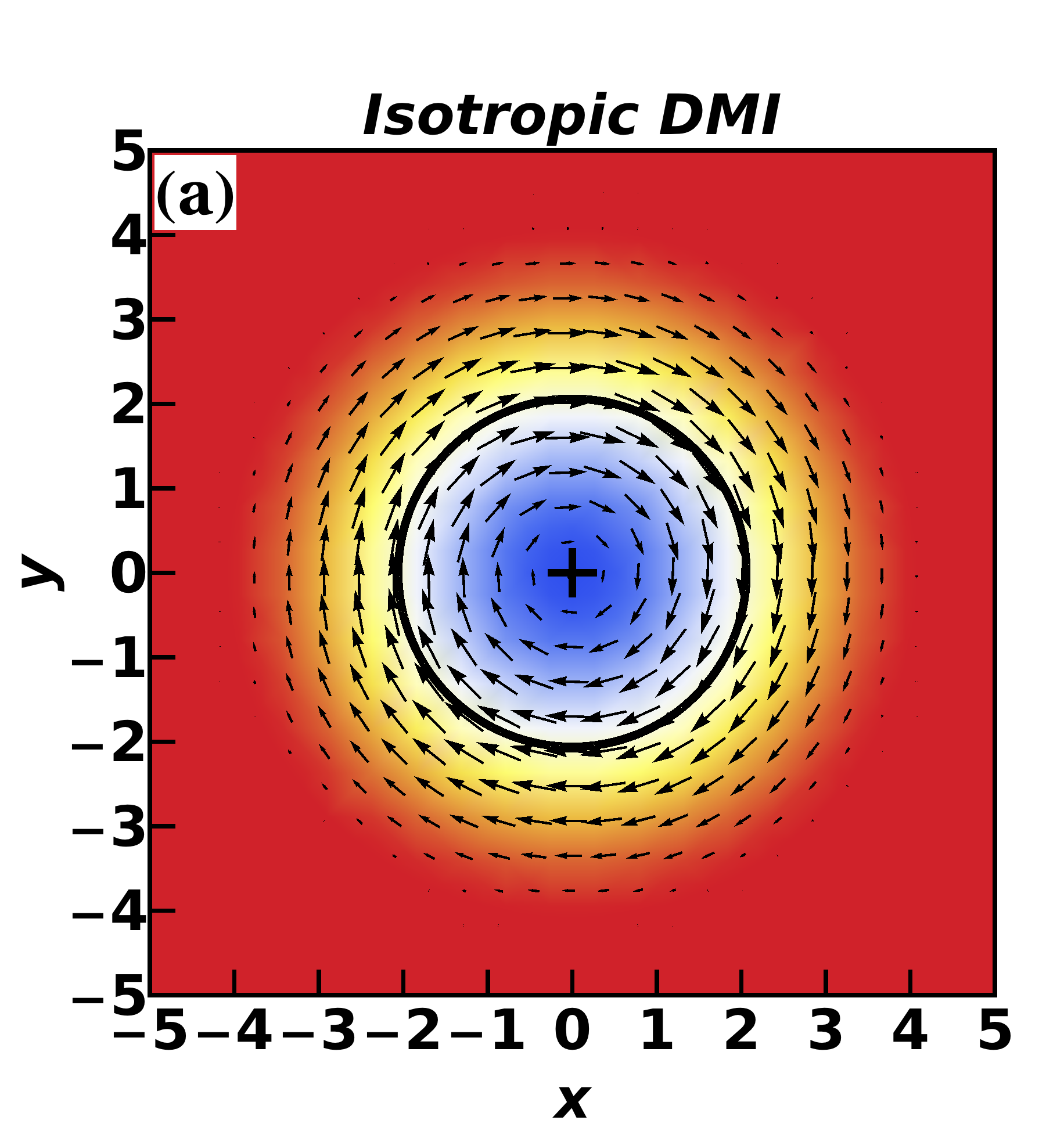}
\label{subfig:fa}
}
\subfigure{
\includegraphics[width=4cm]{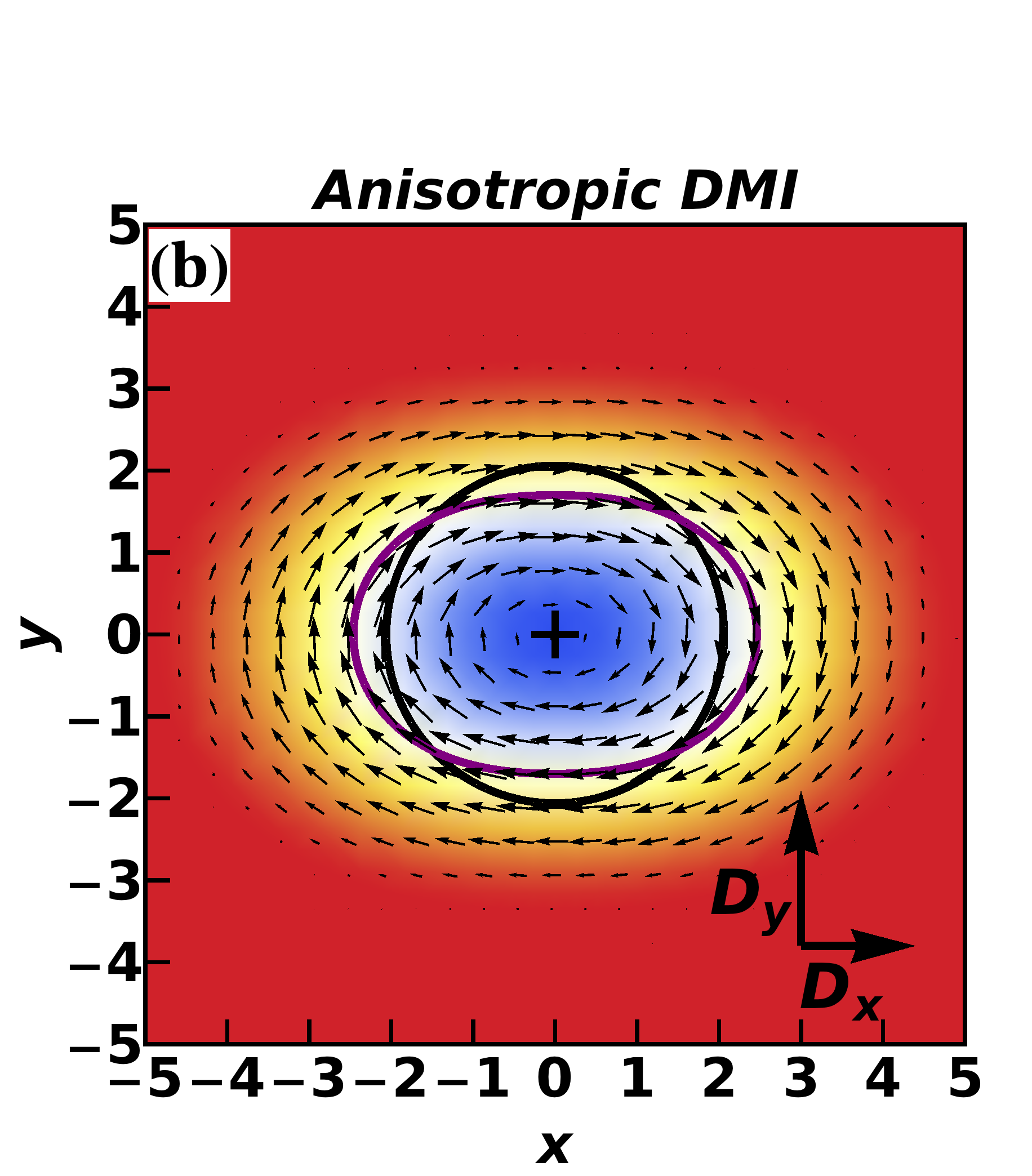}
\label{subfig:fb}
}
\subfigure{
\includegraphics[width=4cm]{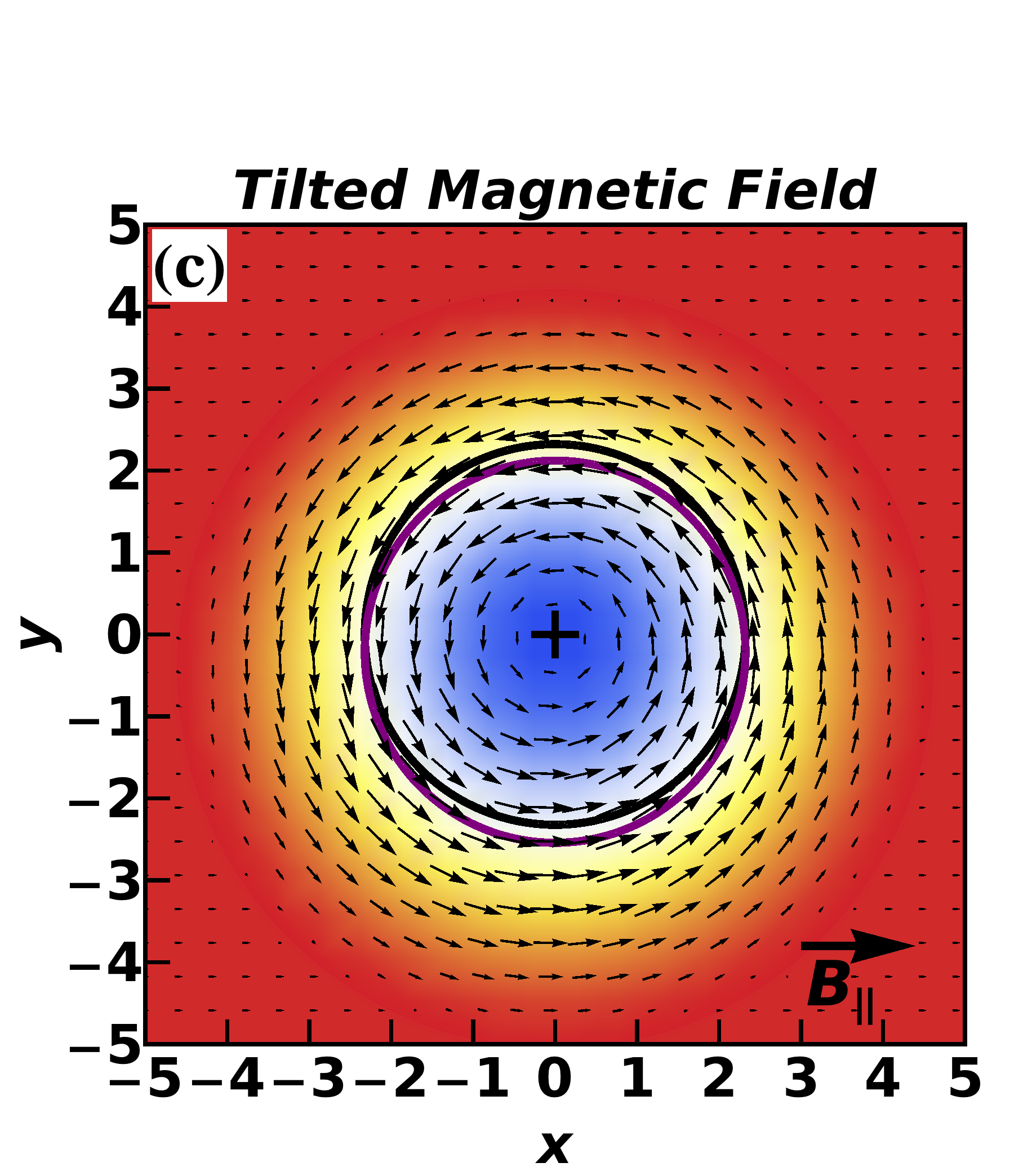}
\label{subfig:fc}
}
\subfigure{
\includegraphics[width=4cm]{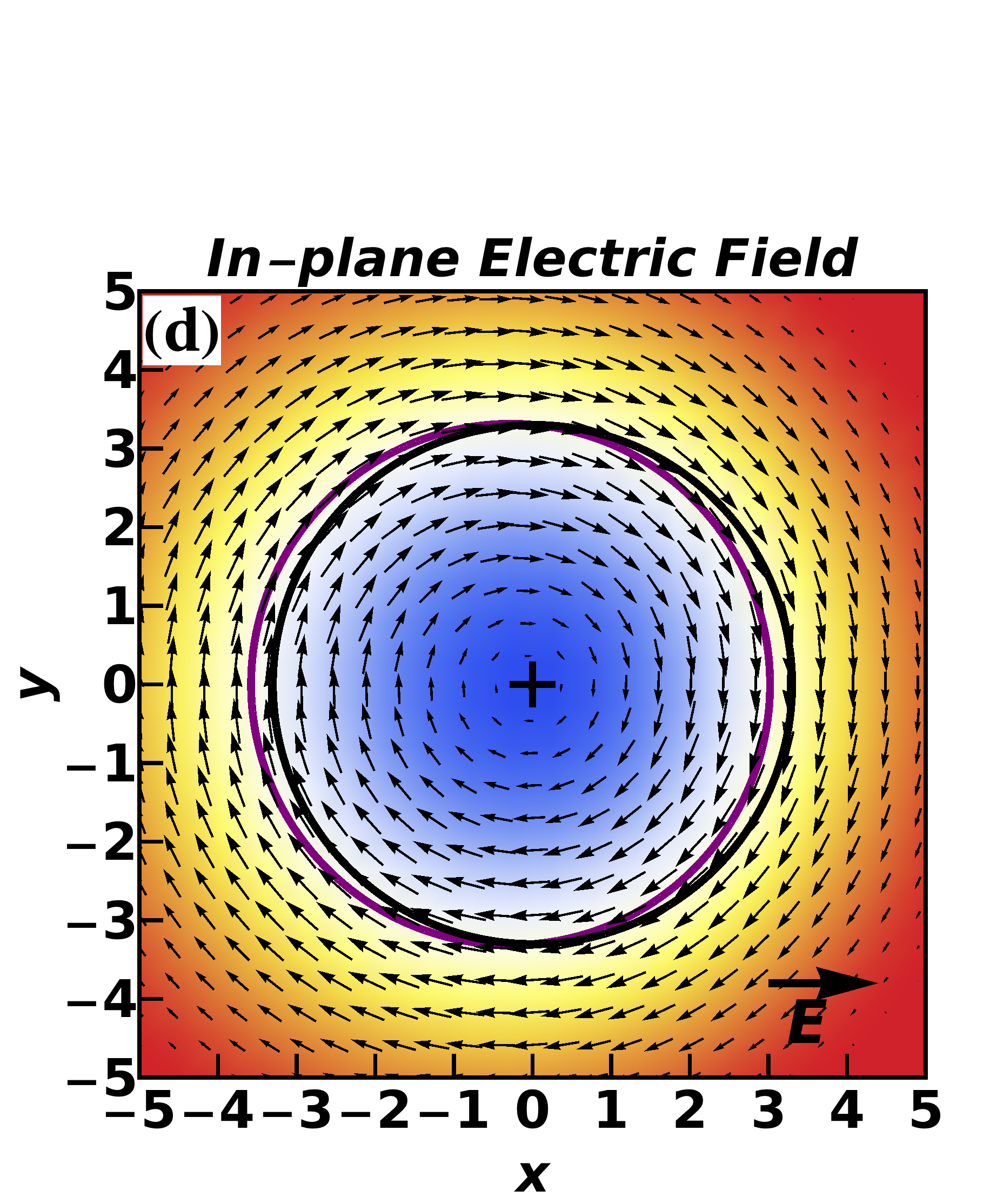}
\label{subfig:fd}
}
\caption{In these pictures we represent the skyrmion deformations through the different perturbations. The curves represents the contours of constant $m_{z}=0$. The black curve is the contour for the circular skyrmion approximation while the purple one is for the distorted skyrmion. The black cross represents the position of the skyrmion center. a) A circular skyrmion for the isotropic DMI ($D/J=0.78$) in a transverse magnetic field ($B/J=0.5$). b) A elliptical skyrmion for the anisotropic DMI model ($D_{y}/J=1$, $D_{x}/J=0.56$ and $B/J=0.5$). c) A distorted skyrmion for a tilted magnetic field ($\theta_{B}=0.18$, $D/J=0.8$ and $B/J=0.45$). d) A distorted skyrmion in an in-plane electric field ($D/J=0.4$, $B/J=0.16$ and $E/J=0.15$).}
\label{fig:tilt}
\end{figure}

In the stability region we determine the shape of the skyrmion by minimizing the energy with respect to $R_{1}$ and $R_{2}$. In Fig. \ref{subfig:fc} we show how the skyrmion shape is affected by an inclination of the magnetic field, where the in-plane component ($\mathbf{B}_{||}$) is in the $\mathbf{\hat{x}}$ ($\phi_{B}=0$) direction.The contours $m_{z}=const.$ are symmetric with respect to reflection along $\mathbf{\hat{y}}$ but not along $\mathbf{\hat{x}}$. In this case the distortion takes place along the $\mathbf{\hat{y}}$ direction, and for arbitrary orientation of $\mathbf{B}_{||}$ the direction of the deformation is determined by Eq. (\ref{eq:betatilted}). This figure can be compared with those presented in results reported recently \cite{LIN_15, WAN_17}.  In these references the authors shown that for a perpendicular magnetic field, the skyrmion is centrosymmetric. As the magnetic field is gradually tilted from the normal vector to the system plane, each skyrmion loses its axis-symmetry. The skyrmion stretches along the direction perpendicular to the in-plane component of the magnetic field ($\mathbf{B}_{||}$). However the shape of the skyrmion is not elliptical. This asymmetry yields a net magnetization moment along the direction of $\mathbf{B}_{||}$ a result confirmed by our calculations.

We compare the previous analytical results with a numerical treatment of the problem for 100 modes expansion. The phase diagram for $\theta_{B}=0.18$ ($\approx 10\degree$) obtained by numerical diagonalization of the Hessian is shown in Fig. \ref{subfig:tilted}. The color background represents the stable (orange) and unstable (blue) regions. The solid black line is the curve $B_{2}^{ZT}$ (obtained by the analytical two modes expansion) which defines the boundary of the stability regions. Regarding the shape of the skyrmion we present the values for $R_{n}$ in the Fig. \ref{subfig:rtilted}. Just the first two modes seems to be relevant in the expansion. These results prove the excellent agreement between the analytical and numerical treatment of the problem.

\subsection{Magnetoelectric coupling}
\label{subsec:magneto}

The compound Cu$_{2}$SeO$_{3}$ \cite{SEKI_12,MOCHI_13} is a multiferroic, where a particular magnetic ordering induces a ferroelectric polarization. It is known that this insulating material hosts Bloch-type skyrmions. Due to its multiferroic nature it is possible to create, manipulate and excite skyrmions by electric fields\cite{MOCHI_13,JSWHITE_12,MOCH_16} which makes it an interesting material for technological applications as memory devices and microwave diodes \cite{MOCHI_13}. In this material the interaction between the electric field and the electric polarization is given by 
\begin{equation}
E_{ME}=-\mathbf{E}\cdot \sum_{i} \mathbf{P}_{i},
\end{equation}
where the local polarization, is related to the spin field in the following way\cite{MOCHI_13, MOCHI_15, MOCHI_16}
\begin{equation}
\mathbf{P}(\mathbf{r})=(S_{z}(\mathbf{r})S_{y}(\mathbf{r}),S_{z}(\mathbf{r})S_{x}(\mathbf{r}),S_{y}(\mathbf{r})S_{x}(\mathbf{r})).
\end{equation}
We are going to analyze the stability of skyrmions under the effect of an external electric field $\mathbf{E}=E(\sin(\gamma)\cos(\alpha),\sin(\gamma)\sin(\alpha),\cos(\gamma))$.
Within our framework the magnetoelectric term is expressed as follows (see Appendix \ref{app:constants}):
\be
E^{ME}=E_{0}^{E}+\sum_{n}^{\infty}L_{n}^{E}R_{n}+\sum_{n,m=1}^{\infty}M_{n m}^{E}R_{n}R_{m}.
\ee

Two relevant cases are considered: 1) transverse electric field ($\gamma=0$) which has effects in the stability of skyrmions, and 2) in-plane electric field ($\gamma=\pi/2$), where the stability of skyrmions is not affected but we find that we can change the shape of the skyrmions.

\subsubsection{Transverse electric field}

In a transverse electric field ($\gamma=0$) the minimum of the energy is reached with $\beta_{n}$ satisfying the equations:
\bea
\nonumber
\beta_{1}&=&\chi_{0}+\frac{\pi}{4},\\
\beta_{n+2}&=&2\chi_{0}+\beta_{n}-\frac{3\pi}{2}.
\label{eq:magnetel}
\eea
It is important to note that a transverse electric field just determines $\beta_{1}$ while $\beta_{2}$ remains as undetermined parameter. The rest of the parameters $\beta_{n}$ are determined through the relation Eq. (\ref{eq:magnetel}).

In this case the magnetoelectric term takes the form:
\be
E^{EP}=\sum_{n,m=1}^{\infty} \tilde{M}_{n m}^{EP}R_{n}R_{m}.
\label{ec:mep}
\ee

We are going to consider just the first four deformations as we did in the previous models. In this approximation we can find analytical expressions for the critical fields. In this case two, critical fields $B_{1}^{MEP}$ and $B_{2}^{MEP}$ defines the boundaries between stable and unstable regions (Fig. \ref{subfig:magnetoelectricperp}). In the limit $E\to0$ we have $B_{1}^{MEP}\to B_{1}^{I}$ and $B_{2}^{MEP}\to B_{2}^{I}$.
Then the phase diagram changes as shown in Fig. \ref{subfig:magnetoelectricperp}. Below $B_{1}^{MEP}$ the instability in the skyrmion shape is driven by the cardioid deformation ($R_{1}$) and, as in the previous cases, below $B_{2}^{MEP}$ the elliptical instability emerges. We see that the critical field $B_{1}^{MEP}$ has a strong dependence on the external electric field. This enables the destruction of a skyrmion by means of an electric field. For example, if the electric field is absent and we create a skyrmion in the point marked with a cross in Fig. \ref{subfig:magnetoelectricperp} we will have a stable single skyrmion. Then, if we turn on the electric field $E/J=0.15$, the previous point becomes an unstable one and the skyrmion could be destroyed. Therefore, our results show that the stability of a skyrmion could be controlled by means of an external electric field. This phenomenon is well known in ferroic systems hosting skyrmions \cite{JSWHITE_12,MOCHI_13}. However our results are not directly comparable to those because the experimental mechanism underlying creation (destruction) of a skyrmion involves localized electric fields.

From the energy expansion Eq. (\ref{ec:mep}) it is easy to see that the shape of the skyrmion will be circular because of the absence of linear terms in $R_{n}$ ($L_{n}^{EP}=0$).

The numerical results (with 100 modes) for the phase diagram (Fig. \ref{subfig:magnetoelectricperp}) are consistent with the analytical results for $B_{1}^{MEP}$ and $B_{2}^{MEP}$.

\subsubsection{In-plane electric field}

In this case $\gamma=\pi/2$ and the values for $\beta$'s are:
\bea
\nonumber
\beta_{1}&=&\frac{3\pi}{2}+\chi_{0}+\alpha,\\ 
\beta_{n+1}&=&\beta_{n}+\chi_{0}+\alpha-\frac{3\pi}{2}.
\label{eq:betamagnetoin}
\eea
The magnetoelectric term takes the form:
\be
E^{EI}=\sum_{n=1}^{\infty} \tilde{L}_{n}^{EI}R_{n}+\sum_{n,m=1}^{\infty} \tilde{M}_{n m}^{EI}R_{n}R_{m}.
\ee

Instead of solving the full series through the recursive previous expression we consider just the first two deformations $R_{1}$ and $R_{2}$. In this approximation the phase diagram is the same as in the isotropic model in the absence of the magnetoelectric term. This result es confirmed numerically as in the previous cases. However the skyrmion shape changes because of the presence of a linear term in $R_{1}$. The most interesting finding is that $R_{1}$ is greater than $R_{2}$ (see Fig. \ref{subfig:rmeperp}). So for small electric fields $R_{1}$ is small and the deformation of the skyrmion can be seen as a translation of the constant $m_z$ contours while the center of the skyrmion remains fixed at the origin of the coordinates. This reveals that the skyrmion shape can be manipulated by the external electric field. The direction of the distortion is determined by the Eqs. (\ref{eq:betamagnetoin}). In particular, in Fig. \ref{subfig:fd}, we show how the skyrmion is distorted by an in-plane electric field in the $\mathbf{\hat{x}}$ ($\alpha=0$) direction and the deformation occurs in the oposite direction (purple contour). This deformation is in accordance with the presence of an electric polarization in the direction of the electric field.   

\section{Conclusions}
\label{sec:conc}

A general spin model where it is possible to find skyrmions contains different kind of interactions as exchange, antisymmetric exchange (Dzyaloshinskii-Moriya), single ion anisotropy, among others. It is also possible to couple the spins to external fields. Each of these interactions determines the phase diagram of the system where a skyrmion could emerge as a metastable state.
We have introduced a method which provides an approach to the systematic study of the stability and skyrmion manipulation. An important feature of the method resides in its fast convergence. In most cases an expansion in a small number of modes is sufficient to capture the physics of the problem, allowing an analytical approach. In more complex cases, requiring a large number of modes, a numerical treatment of the problem is still possible.\\
The method reproduces satisfactorily the available experimental and numerical results which we summarize below.

For the isotropic model: we find circular skyrmions and its stability region in the phase diagram, and the critical field for the transition SkX $\to$ H which corresponds to the emergence of elliptical deformations of the skyrmions.

For the anisotropic model we find the stability region were elliptical skyrmions are stabilized. In view of our findings we see that the elliptical shape of the skyrmions is due to the presence of an anisotropic DMI, indeed the elliptical deformation is proportional to the anisotropy: $R_{2}\varpropto D^{-}$. As we have shown for this model, in the low anisotropy region the instability is elliptical while in the high anisotropy limit is triangular. 

For the isotropic model in a tilted magnetic field we find that the phase diagram remains almost unchanged with respect to the transverse magnetic field case, except for the increase of the critical field as the tilting angle increases. The most important consequence is a slight reduction of the skyrmion stability region. The shape of the skyrmions changes losing its axial symmetry. This fact is consistent with both experimental \cite{WAN_17} and numerical \cite{LIN_15} results.

Finally we have studied the effect of an external electric field in the stability of a skyrmion. When a transverse  electric field is applied the skyrmion can become unstable through a cardioid instability, leading to a mechanism for the destruction of skyrmions. On the other hand, an in-plane electric field could serve as an element to control the shape of the skyrmion core. To the best of our knowledge there are no experimental results for a longitudinal electric field but for the sake of completeness we present our predictions for that case. We hope that this will prompt experiments dealing with this situation.

\appendix
\section{Condition on $\chi$}
\label{app:chi}

In order to clarify the origin of the expression for $\chi(\phi)$ we derive the condition in Eq. (\ref{ec:condtchernysov}).
We start by writing the tangent ($\mathbf{v}_{t}$) and perpendicular ($\mathbf{v}_{p}$) vectors to the curve $R(\phi)$, 
\bea
\mathbf{v}_{t}=(\frac{\partial x}{\partial \phi},\frac{\partial y}{\partial \phi}),\\
\mathbf{v}_{p}=(\frac{\partial y}{\partial \phi},-\frac{\partial x}{\partial \phi}),\\
\eea
where $x=R(\phi)\cos(\phi)$ and $y=R(\phi)\sin(\phi)$. We write $\mathbf{v}_t=v(\phi)(-\sin(\psi),\cos(\psi))$ and $\mathbf{v}_p=v(\phi)(\cos(\psi),\sin(\psi))$ with $\psi=\phi+\alpha(\phi)$. If the curve were a circle then the normal vector is given by $\psi=\phi$. We want to find $\alpha(\phi)$ in terms of $R(\phi)$ so we start with the equations:

\bea
(\frac{\partial x}{\partial \phi},\frac{\partial y}{\partial \phi})=v(\phi)(-\sin(\psi),\cos(\psi)),\nonumber \\
(\frac{\partial y}{\partial \phi},-\frac{\partial x}{\partial \phi})=v(\phi)(\cos(\psi),\sin(\psi)).
\label{eq:conc}
\eea
We are considering curves where the normal vector can be defined, so we require a single-valued function $R(\phi)$ and $v(\phi)\neq0$. Then from Eqs. (\ref{eq:conc}) we find

\be
\arctan(\alpha)=\frac{R'(\phi)}{R(\phi)}.
\ee
We are interested in stability analysis which requires the expansion of the energy up to second order in deformations, so we expand the left hand side up to second order in $\alpha$ and right hand side of the previous equation up to second order in $R_{n}$ 

\bea
\arctan(\alpha)\approx\alpha+\mathcal{O}(\alpha^{3}),\\
\frac{R'(\phi)}{R(\phi)}\approx-\frac{\delta R(\phi)'}{R_{0}}\left[1-\frac{\delta R(\phi)}{R_{0}}\right].
\eea
Finally we solve for $\alpha(\phi)$ and we find
\be
\alpha=-\frac{\delta R(\phi)'}{R_{0}}+\frac{\delta R(\phi)}{R_{0}}\frac{\delta R(\phi)'}{R_{0}}.
\ee
Then we write $\chi(\phi)=\chi_{0}+\alpha(\phi)$. If $\chi_{0}=0$ the vector $(\cos(\phi+\chi(\phi)),\sin(\phi+\chi(\phi)))$ is perpendicular to the curve $R(\phi)$ and if $\chi_{0}=\pm\pi/2$ the vector is tangent to the curve. As a final comment we mention that for $R(\phi)$ to be smooth and single-valued it is sufficient $\alpha\ll1$. For this condition to be fulfilled it is sufficient the condition $\sum_{n=1}^{\infty}n |R_{n}|\ll R_{0}$ which, in turn, implies $|\delta R(\phi)|\ll R_0$.

\section{Matrix $\mathbf{M}$ and vector $\mathbf{L}$ for each model}
\label{app:constants}

Since the function $R(\phi)$ represents a scale change of the coordinates, the radial integrals can be easily computed through the substitution $u=r/R(\phi)$. These integrals are constants independent of the parameters of the model (except for $\tilde{\Lambda}_{3}$ which depends on $\theta_{B}$, the inclination angle of the external field) and appear in all our calculations.We list below the set of constants that appear throughout all of the work:

\bea
\nonumber
\Lambda_{1}&=&\int_{0}^{1} u (f \ensuremath{'} (u))^{2} du = \frac{\pi^{2}}{2},\\
\nonumber
\Lambda_{2}&=&\int_{0}^{1} \frac{\sin^{2}[f(u)]}{u} du =\\
\nonumber
&=&\frac{\gamma-CosIntegral[2\pi]+\log[2\pi]}{2},\\
\nonumber
\tilde{\Lambda}_{3}&=&\int_{0}^{1} u \cos(\theta_B)\cos[f(u)] du=\\
\nonumber
&=&\frac{2\cos(\theta_{B})}{\pi^{2}}-\frac{1}{2},\\
\nonumber
\nonumber
\Lambda_{4}&=&\int_{0}^{1} u f \ensuremath{'} (u) du = -\frac{\pi}{2},\\
\nonumber
\tilde{\Lambda}_{5}&=&\frac{\sin(\theta_{B})}{\pi},
\eea
where $\gamma$ is the Euler's constant ($\gamma\approx 0.577216$). For $\theta_{B}=0$ we have $\tilde{\Lambda}_{3}=\Lambda_{3}$ and $\tilde{\Lambda}_{5}=0$. 

The energy, in a specific model, is expressed as a sum of the contribution from each interaction.\\
Thus for anisotropic DMI model in a transverse magnetic field we have:
\bea
E_{0}^{A}&=&E_{0}^{J}+E_{0}^{DA}+E_{0}^{Z},\\
L_{n}^{A}&=&L_{n}^{J}+L_{n}^{DA}+L_{n}^{Z},\\
M_{n m}^{A}&=&M_{n m}^{J}+M_{n m}^{DA}+M_{n m}^{Z}.
\eea
For the isotropic DMI model in a tilted magnetic field we have:
\bea
E_{0}^{T}&=&E_{0}^{J}+E_{0}^{DI}+E_{0}^{ZT},\\
L_{n}^{T}&=&L_{n}^{J}+L_{n}^{DI}+L_{n}^{ZT},\\
M_{n m}^{T}&=&M_{n m}^{J}+M_{n m}^{DI}+M_{n m}^{ZT}.
\eea
With the addition of the magneto-electric term to the isotropic model (in transverse magnetic field) we have:
\bea
E_{0}^{ME}&=&E_{0}^{J}+E_{0}^{DI}+E_{0}^{Z}+E_{0}^{E},\\
L_{n}^{ME}&=&L_{n}^{J}+L_{n}^{DI}+L_{n}^{Z}+L_{n}^{E},\\
M_{n m}^{ME}&=&M_{n m}^{J}+M_{n m}^{DI}+M_{n m}^{Z}+M_{n m}^{E}.
\eea
\begin{widetext}
We write here the respective expressions for $E_{0}$, $L_{n}$ and $M_{n m}$ for each interaction term. For the exchange term: 

\bea
E_{0}^{J}&=&J\pi(\Lambda_{1}+\Lambda_{2}),\\ \nonumber
L_{n}^{J}&=&0,\\ \nonumber
M_{n m}^{J}&=&\frac{J\pi}{2 R_{0}^{2}}(\Lambda_{1}n^{2}+\Lambda_{2}n^{4})\delta_{n m}.
\eea
For the anisotropic DMI (the isotropic case is the particular case of this one in which $D_{x}=D_{y}=D$):

\bea
E_{0}^{DA}&=&-\Lambda_{4}D^{+}\pi R_{0}\sin(\chi_{0}),\\ \nonumber
L_{n}^{DA}&=&\Lambda_{4}D^{-}\pi\frac{3}{2}\sin(\beta_{2}-\chi_{0})\delta_{n 2},\\ \nonumber
M_{n m}^{DA}&=&\Lambda_{4}D^{-}\pi\frac{3}{8R_{0}}\sin(\chi_{0}-2\beta_{1})\delta_{n 1}\delta_{m 1}-\Lambda_{4}D^{+}\pi\frac{\pi}{4R_{0}}\sin(\chi_{0})n^{2}\delta_{n m}-\\ \nonumber
&-&\frac{3\Lambda_{4}D^{-}\pi}{8R_{0}}\left[n(n+2)\sin(\beta_{n}-\beta_{n+2}+\chi_{0})\delta_{m n+2}+m(m+2)\sin(\beta_{m}-\beta_{m+2}+\chi_{0})\delta_{n m+2}\right].
\eea
The Zeemann term for a tilted magnetic field:

\bea
E_{0}^{ZT}&=&-B\tilde{\Lambda}_{3}2\pi R_{0}^{2},\\ \nonumber
L_{n}^{ZT}&=&-B\tilde{\Lambda}_{5}\pi R_{0}\cos(\phi_{B}+\beta_{1}-\chi_{0})\delta_{1 n},\\ \nonumber
M_{n m}^{ZT}&=&-B\tilde{\Lambda}_{3}\pi\delta_{n m}-\\ \nonumber
&-&B\tilde{\Lambda}_{5}\frac{\pi}{4}\left[(\frac{3}{2}+n(n+1))\cos(\beta_{n+1}-\beta_{n}+\phi_{B}-\chi_{0})\delta_{m n+1}+(\frac{3}{2}+m(m+1))\cos(\beta_{m+1}-\beta_{m}+\phi_{B}-\chi_{0})\delta_{n m+1}\right].
\eea
The transverse field configuration ($E_{0}^{Z}$, $L_{n}^{Z}$ and $M_{n m}^{Z}$) is obtained by setting $\theta_{B}=0$.

Finally the magneto-electric term contributes with:
\bea
E_{0}^{E}&=&0,\\ \nonumber
L_{n}^{E}&=& E \sin(\gamma)\frac{R_{0}}{4}\sin(\beta_{1}-\chi_{0}-\alpha),\\ \nonumber
M_{n m}^{E}&=&- E \frac{3\pi}{32}\cos(\gamma)\sin(2\beta_{1}-2\chi_{0})\delta_{n 1}\delta_{m 1}-\\ \nonumber
&-&\frac{E \sin(\gamma)}{8}\left[(\frac{1-2n}{4})\sin(\beta_{n}-\beta_{n+1}+\chi_{0}+\alpha)\delta_{m n+1}+(\frac{1-2m}{4})\sin(\beta_{m}-\beta_{m+1}+\chi_{0}+\alpha)\delta_{n m+1}\right]+\\ \nonumber
&+&\frac{E \pi}{16}\cos(\gamma)\left[\sin(2\chi_{0}+\beta_{n}-\beta_{n+2})(1+2n(n+2))\delta_{m n+2}+\sin(2\chi_{0}+\beta_{m}-\beta_{m+2})(1+2m(m+2))\delta_{n m+2}\right].
\eea
In each case the equilibrium radius ($R_0$) and helicity ($\chi_{0}$), as well as the $\beta_{n}$ parameters, must be calculated for the corresponding model as explained in the main text of the work. 

\end{widetext}
\section{Relevance of the discontinuity}
\label{app:discon}
To estimate the contribution of the discontinuity to the energy we suppose that the spin field varies from the skyrmion boundary (at $r=R$) and the variation occurs in a length of the order of the lattice parameter $a$ so the field polarized state is reached at $r=R+a$. This variation takes place in the crown of internal radii $R$ and external radii $R+a$. We express the spin field as in \ref{ec:compmag} where the function $\Theta$ and $\Phi$ are given by:

\bea
\Theta(r)=\frac{\theta_{B}}{a}(r-R), & R\leq r\leq R+a, \\
\Phi(r)=\frac{\phi_{B}}{a}(r-R), & R\leq r\leq R+a.
\eea

We evaluate the energy (which we shall call $E_{D}$) through the previous approximation and subtract from it the energy of the crown with their spin aligned with the magnetic field ($E_{FP}$). Since we are interested in small angles $\theta_{B}$ we expand the previous result up to second order in $\theta_{B}$ to obtain:
\begin{widetext}
\bea
& &E_{D} - E_{FP} \approx \\
\nonumber
&\approx& \left\lbrace B \left[ 9 a ^{2} + 16 a R_{0} - 48 \frac{a^{2}}{\phi_{B}^{2}} - 24\frac{a R_{0}}{\phi_{B}^{2}}  + 24 \frac{a R_{0}\cos[\phi_{B}]}{\phi_{B}^{2}} + 48 \frac{a^{2} \sin[\phi_{B}]}{\phi_{B}^{3}}\right] + J \left[6 +  12 \frac{R_{0}}{a}  + 3 \phi_{B}^{2} + 4\frac{R_{0} \phi_{B}^{2}}{a}\right]\right\rbrace\frac{\theta_{B}^{2}}{12\pi}+\\
\nonumber
&+& \mathcal{O}(\theta_{B}^{4}),
\eea
We can see that this difference is of order $\theta_{B}^{2}$ as mentioned before.
\end{widetext}

\bibliography{references.bib}

\newcommand{\npb}{Nucl. Phys. B}\newcommand{\adv}{Adv.
  Phys.}\newcommand{\epl}{Europhys. Lett.}
\begin{thebibliography}{40}
\expandafter\ifx\csname natexlab\endcsname\relax\def\natexlab#1{#1}\fi
\expandafter\ifx\csname bibnamefont\endcsname\relax
  \def\bibnamefont#1{#1}\fi
\expandafter\ifx\csname bibfnamefont\endcsname\relax
  \def\bibfnamefont#1{#1}\fi
\expandafter\ifx\csname citenamefont\endcsname\relax
  \def\citenamefont#1{#1}\fi
\expandafter\ifx\csname url\endcsname\relax
  \def\url#1{\texttt{#1}}\fi
\expandafter\ifx\csname urlprefix\endcsname\relax\def\urlprefix{URL }\fi
\providecommand{\bibinfo}[2]{#2}
\providecommand{\eprint}[2][]{\url{#2}}

\bibitem[{\citenamefont{Nagaosa and Tokura}(2013)}]{NagaosaTokura2013}
\bibinfo{author}{\bibfnamefont{N.}~\bibnamefont{Nagaosa}} \bibnamefont{and}
  \bibinfo{author}{\bibfnamefont{Y.}~\bibnamefont{Tokura}},
  \bibinfo{journal}{Nature nanotechnology} \textbf{\bibinfo{volume}{8}},
  \bibinfo{pages}{899} (\bibinfo{year}{2013}).

\bibitem[{\citenamefont{M{\"u}hlbauer et~al.}(2009)\citenamefont{M{\"u}hlbauer,
  Binz, Jonietz, Pfleiderer, Rosch, Neubauer, Georgii, and B{\"o}ni}}]{MBJ_09}
\bibinfo{author}{\bibfnamefont{S.}~\bibnamefont{M{\"u}hlbauer}},
  \bibinfo{author}{\bibfnamefont{B.}~\bibnamefont{Binz}},
  \bibinfo{author}{\bibfnamefont{F.}~\bibnamefont{Jonietz}},
  \bibinfo{author}{\bibfnamefont{C.}~\bibnamefont{Pfleiderer}},
  \bibinfo{author}{\bibfnamefont{A.}~\bibnamefont{Rosch}},
  \bibinfo{author}{\bibfnamefont{A.}~\bibnamefont{Neubauer}},
  \bibinfo{author}{\bibfnamefont{R.}~\bibnamefont{Georgii}}, \bibnamefont{and}
  \bibinfo{author}{\bibfnamefont{P.}~\bibnamefont{B{\"o}ni}},
  \bibinfo{journal}{Science} \textbf{\bibinfo{volume}{323}},
  \bibinfo{pages}{915} (\bibinfo{year}{2009}).

\bibitem[{\citenamefont{Ishikawa and Arai}(1984)}]{IA_84}
\bibinfo{author}{\bibfnamefont{Y.}~\bibnamefont{Ishikawa}} \bibnamefont{and}
  \bibinfo{author}{\bibfnamefont{M.}~\bibnamefont{Arai}},
  \bibinfo{journal}{Journal of the Physical Society of Japan}
  \textbf{\bibinfo{volume}{53}}, \bibinfo{pages}{2726} (\bibinfo{year}{1984}).

\bibitem[{\citenamefont{Lebech et~al.}(1995)\citenamefont{Lebech, Harris,
  Pedersen, Mortensen, Gregory, Bernhoeft, Jermy, and Brown}}]{LHP_95}
\bibinfo{author}{\bibfnamefont{B.}~\bibnamefont{Lebech}},
  \bibinfo{author}{\bibfnamefont{P.}~\bibnamefont{Harris}},
  \bibinfo{author}{\bibfnamefont{J.~S.} \bibnamefont{Pedersen}},
  \bibinfo{author}{\bibfnamefont{K.}~\bibnamefont{Mortensen}},
  \bibinfo{author}{\bibfnamefont{C.}~\bibnamefont{Gregory}},
  \bibinfo{author}{\bibfnamefont{N.}~\bibnamefont{Bernhoeft}},
  \bibinfo{author}{\bibfnamefont{M.}~\bibnamefont{Jermy}}, \bibnamefont{and}
  \bibinfo{author}{\bibfnamefont{S.}~\bibnamefont{Brown}},
  \bibinfo{journal}{Journal of magnetism and magnetic materials}
  \textbf{\bibinfo{volume}{140}}, \bibinfo{pages}{119} (\bibinfo{year}{1995}).

\bibitem[{\citenamefont{Shibata et~al.}(2013)\citenamefont{Shibata, Yu, Hara,
  Morikawa, Kanazawa, Kimoto, Ishiwata, Matsui, and Tokura}}]{SYH_13}
\bibinfo{author}{\bibfnamefont{K.}~\bibnamefont{Shibata}},
  \bibinfo{author}{\bibfnamefont{X.}~\bibnamefont{Yu}},
  \bibinfo{author}{\bibfnamefont{T.}~\bibnamefont{Hara}},
  \bibinfo{author}{\bibfnamefont{D.}~\bibnamefont{Morikawa}},
  \bibinfo{author}{\bibfnamefont{N.}~\bibnamefont{Kanazawa}},
  \bibinfo{author}{\bibfnamefont{K.}~\bibnamefont{Kimoto}},
  \bibinfo{author}{\bibfnamefont{S.}~\bibnamefont{Ishiwata}},
  \bibinfo{author}{\bibfnamefont{Y.}~\bibnamefont{Matsui}}, \bibnamefont{and}
  \bibinfo{author}{\bibfnamefont{Y.}~\bibnamefont{Tokura}},
  \bibinfo{journal}{Nature nanotechnology} \textbf{\bibinfo{volume}{8}},
  \bibinfo{pages}{723} (\bibinfo{year}{2013}).

\bibitem[{\citenamefont{Lebech et~al.}(1989)\citenamefont{Lebech, Bernhard, and
  Freltoft}}]{LBF_89}
\bibinfo{author}{\bibfnamefont{B.}~\bibnamefont{Lebech}},
  \bibinfo{author}{\bibfnamefont{J.}~\bibnamefont{Bernhard}}, \bibnamefont{and}
  \bibinfo{author}{\bibfnamefont{T.}~\bibnamefont{Freltoft}},
  \bibinfo{journal}{Journal of Physics: Condensed Matter}
  \textbf{\bibinfo{volume}{1}}, \bibinfo{pages}{6105} (\bibinfo{year}{1989}).

\bibitem[{\citenamefont{Uchida et~al.}(2008)\citenamefont{Uchida, Nagaosa, He,
  Kaneko, Iguchi, Matsui, and Tokura}}]{UNH_08}
\bibinfo{author}{\bibfnamefont{M.}~\bibnamefont{Uchida}},
  \bibinfo{author}{\bibfnamefont{N.}~\bibnamefont{Nagaosa}},
  \bibinfo{author}{\bibfnamefont{J.}~\bibnamefont{He}},
  \bibinfo{author}{\bibfnamefont{Y.}~\bibnamefont{Kaneko}},
  \bibinfo{author}{\bibfnamefont{S.}~\bibnamefont{Iguchi}},
  \bibinfo{author}{\bibfnamefont{Y.}~\bibnamefont{Matsui}}, \bibnamefont{and}
  \bibinfo{author}{\bibfnamefont{Y.}~\bibnamefont{Tokura}},
  \bibinfo{journal}{Physical Review B} \textbf{\bibinfo{volume}{77}},
  \bibinfo{pages}{184402} (\bibinfo{year}{2008}).

\bibitem[{\citenamefont{Yu et~al.}(2011)\citenamefont{Yu, Kanazawa, Onose,
  Kimoto, Zhang, Ishiwata, Matsui, and Tokura}}]{YKO_11}
\bibinfo{author}{\bibfnamefont{X.}~\bibnamefont{Yu}},
  \bibinfo{author}{\bibfnamefont{N.}~\bibnamefont{Kanazawa}},
  \bibinfo{author}{\bibfnamefont{Y.}~\bibnamefont{Onose}},
  \bibinfo{author}{\bibfnamefont{K.}~\bibnamefont{Kimoto}},
  \bibinfo{author}{\bibfnamefont{W.}~\bibnamefont{Zhang}},
  \bibinfo{author}{\bibfnamefont{S.}~\bibnamefont{Ishiwata}},
  \bibinfo{author}{\bibfnamefont{Y.}~\bibnamefont{Matsui}}, \bibnamefont{and}
  \bibinfo{author}{\bibfnamefont{Y.}~\bibnamefont{Tokura}},
  \bibinfo{journal}{Nature materials} \textbf{\bibinfo{volume}{10}},
  \bibinfo{pages}{106} (\bibinfo{year}{2011}).

\bibitem[{\citenamefont{Wilhelm et~al.}(2011)\citenamefont{Wilhelm, Baenitz,
  Schmidt, R{\"o}{\ss}ler, Leonov, and Bogdanov}}]{WBS_11}
\bibinfo{author}{\bibfnamefont{H.}~\bibnamefont{Wilhelm}},
  \bibinfo{author}{\bibfnamefont{M.}~\bibnamefont{Baenitz}},
  \bibinfo{author}{\bibfnamefont{M.}~\bibnamefont{Schmidt}},
  \bibinfo{author}{\bibfnamefont{U.}~\bibnamefont{R{\"o}{\ss}ler}},
  \bibinfo{author}{\bibfnamefont{A.}~\bibnamefont{Leonov}}, \bibnamefont{and}
  \bibinfo{author}{\bibfnamefont{A.}~\bibnamefont{Bogdanov}},
  \bibinfo{journal}{Physical review letters} \textbf{\bibinfo{volume}{107}},
  \bibinfo{pages}{127203} (\bibinfo{year}{2011}).

\bibitem[{\citenamefont{Beille et~al.}(1983)\citenamefont{Beille, Voiron, and
  Roth}}]{BVR_83}
\bibinfo{author}{\bibfnamefont{J.}~\bibnamefont{Beille}},
  \bibinfo{author}{\bibfnamefont{J.}~\bibnamefont{Voiron}}, \bibnamefont{and}
  \bibinfo{author}{\bibfnamefont{M.}~\bibnamefont{Roth}},
  \bibinfo{journal}{Solid state communications} \textbf{\bibinfo{volume}{47}},
  \bibinfo{pages}{399} (\bibinfo{year}{1983}).

\bibitem[{\citenamefont{Grigoriev et~al.}(2007)\citenamefont{Grigoriev,
  Dyadkin, Menzel, Schoenes, Chetverikov, Okorokov, Eckerlebe, and
  Maleyev}}]{GDM_07}
\bibinfo{author}{\bibfnamefont{S.}~\bibnamefont{Grigoriev}},
  \bibinfo{author}{\bibfnamefont{V.}~\bibnamefont{Dyadkin}},
  \bibinfo{author}{\bibfnamefont{D.}~\bibnamefont{Menzel}},
  \bibinfo{author}{\bibfnamefont{J.}~\bibnamefont{Schoenes}},
  \bibinfo{author}{\bibfnamefont{Y.~O.} \bibnamefont{Chetverikov}},
  \bibinfo{author}{\bibfnamefont{A.}~\bibnamefont{Okorokov}},
  \bibinfo{author}{\bibfnamefont{H.}~\bibnamefont{Eckerlebe}},
  \bibnamefont{and} \bibinfo{author}{\bibfnamefont{S.}~\bibnamefont{Maleyev}},
  \bibinfo{journal}{Physical Review B} \textbf{\bibinfo{volume}{76}},
  \bibinfo{pages}{224424} (\bibinfo{year}{2007}).

\bibitem[{\citenamefont{Grigoriev et~al.}(2009)\citenamefont{Grigoriev,
  Chernyshov, Dyadkin, Dmitriev, Maleyev, Moskvin, Menzel, Schoenes, and
  Eckerlebe}}]{GCD_09}
\bibinfo{author}{\bibfnamefont{S.}~\bibnamefont{Grigoriev}},
  \bibinfo{author}{\bibfnamefont{D.}~\bibnamefont{Chernyshov}},
  \bibinfo{author}{\bibfnamefont{V.}~\bibnamefont{Dyadkin}},
  \bibinfo{author}{\bibfnamefont{V.}~\bibnamefont{Dmitriev}},
  \bibinfo{author}{\bibfnamefont{S.}~\bibnamefont{Maleyev}},
  \bibinfo{author}{\bibfnamefont{E.}~\bibnamefont{Moskvin}},
  \bibinfo{author}{\bibfnamefont{D.}~\bibnamefont{Menzel}},
  \bibinfo{author}{\bibfnamefont{J.}~\bibnamefont{Schoenes}}, \bibnamefont{and}
  \bibinfo{author}{\bibfnamefont{H.}~\bibnamefont{Eckerlebe}},
  \bibinfo{journal}{Physical Review Letters} \textbf{\bibinfo{volume}{102}},
  \bibinfo{pages}{037204} (\bibinfo{year}{2009}).

\bibitem[{\citenamefont{Onose et~al.}(2005)\citenamefont{Onose, Takeshita,
  Terakura, Takagi, and Tokura}}]{OTT_05}
\bibinfo{author}{\bibfnamefont{Y.}~\bibnamefont{Onose}},
  \bibinfo{author}{\bibfnamefont{N.}~\bibnamefont{Takeshita}},
  \bibinfo{author}{\bibfnamefont{C.}~\bibnamefont{Terakura}},
  \bibinfo{author}{\bibfnamefont{H.}~\bibnamefont{Takagi}}, \bibnamefont{and}
  \bibinfo{author}{\bibfnamefont{Y.}~\bibnamefont{Tokura}},
  \bibinfo{journal}{Physical Review B} \textbf{\bibinfo{volume}{72}},
  \bibinfo{pages}{224431} (\bibinfo{year}{2005}).

\bibitem[{\citenamefont{K{\'e}zsm{\'a}rki
  et~al.}(2015)\citenamefont{K{\'e}zsm{\'a}rki, Bord{\'a}cs, Milde, Neuber,
  Eng, White, R{\o}nnow, Dewhurst, Mochizuki, Yanai et~al.}}]{KEZ_15}
\bibinfo{author}{\bibfnamefont{I.}~\bibnamefont{K{\'e}zsm{\'a}rki}},
  \bibinfo{author}{\bibfnamefont{S.}~\bibnamefont{Bord{\'a}cs}},
  \bibinfo{author}{\bibfnamefont{P.}~\bibnamefont{Milde}},
  \bibinfo{author}{\bibfnamefont{E.}~\bibnamefont{Neuber}},
  \bibinfo{author}{\bibfnamefont{L.}~\bibnamefont{Eng}},
  \bibinfo{author}{\bibfnamefont{J.}~\bibnamefont{White}},
  \bibinfo{author}{\bibfnamefont{H.~M.} \bibnamefont{R{\o}nnow}},
  \bibinfo{author}{\bibfnamefont{C.}~\bibnamefont{Dewhurst}},
  \bibinfo{author}{\bibfnamefont{M.}~\bibnamefont{Mochizuki}},
  \bibinfo{author}{\bibfnamefont{K.}~\bibnamefont{Yanai}},
  \bibnamefont{et~al.}, \bibinfo{journal}{Nature materials}
  \textbf{\bibinfo{volume}{14}}, \bibinfo{pages}{1116} (\bibinfo{year}{2015}).

\bibitem[{\citenamefont{Seki et~al.}(2012)\citenamefont{Seki, Yu, Ishiwata, and
  Tokura}}]{SEKI_12}
\bibinfo{author}{\bibfnamefont{S.}~\bibnamefont{Seki}},
  \bibinfo{author}{\bibfnamefont{X.}~\bibnamefont{Yu}},
  \bibinfo{author}{\bibfnamefont{S.}~\bibnamefont{Ishiwata}}, \bibnamefont{and}
  \bibinfo{author}{\bibfnamefont{Y.}~\bibnamefont{Tokura}},
  \bibinfo{journal}{Science} \textbf{\bibinfo{volume}{336}},
  \bibinfo{pages}{198} (\bibinfo{year}{2012}).

\bibitem[{\citenamefont{Adams et~al.}(2012)\citenamefont{Adams, Chacon, Wagner,
  Bauer, Brandl, Pedersen, Berger, Lemmens, and Pfleiderer}}]{ADAM_12}
\bibinfo{author}{\bibfnamefont{T.}~\bibnamefont{Adams}},
  \bibinfo{author}{\bibfnamefont{A.}~\bibnamefont{Chacon}},
  \bibinfo{author}{\bibfnamefont{M.}~\bibnamefont{Wagner}},
  \bibinfo{author}{\bibfnamefont{A.}~\bibnamefont{Bauer}},
  \bibinfo{author}{\bibfnamefont{G.}~\bibnamefont{Brandl}},
  \bibinfo{author}{\bibfnamefont{B.}~\bibnamefont{Pedersen}},
  \bibinfo{author}{\bibfnamefont{H.}~\bibnamefont{Berger}},
  \bibinfo{author}{\bibfnamefont{P.}~\bibnamefont{Lemmens}}, \bibnamefont{and}
  \bibinfo{author}{\bibfnamefont{C.}~\bibnamefont{Pfleiderer}},
  \bibinfo{journal}{Physical review letters} \textbf{\bibinfo{volume}{108}},
  \bibinfo{pages}{237204} (\bibinfo{year}{2012}).

\bibitem[{\citenamefont{Bogdanov and Yablonskii}(1989)}]{BY_89}
\bibinfo{author}{\bibfnamefont{A.}~\bibnamefont{Bogdanov}} \bibnamefont{and}
  \bibinfo{author}{\bibfnamefont{D.}~\bibnamefont{Yablonskii}},
  \bibinfo{journal}{Zh. Eksp. Teor. Fiz} \textbf{\bibinfo{volume}{95}},
  \bibinfo{pages}{182} (\bibinfo{year}{1989}).

\bibitem[{\citenamefont{Bogdanov and Hubert}(1994{\natexlab{a}})}]{BH_94}
\bibinfo{author}{\bibfnamefont{A.}~\bibnamefont{Bogdanov}} \bibnamefont{and}
  \bibinfo{author}{\bibfnamefont{A.}~\bibnamefont{Hubert}},
  \bibinfo{journal}{Journal of magnetism and magnetic materials}
  \textbf{\bibinfo{volume}{138}}, \bibinfo{pages}{255}
  (\bibinfo{year}{1994}{\natexlab{a}}).

\bibitem[{\citenamefont{Bogdanov and Hubert}(1994{\natexlab{b}})}]{BH2_94}
\bibinfo{author}{\bibfnamefont{A.}~\bibnamefont{Bogdanov}} \bibnamefont{and}
  \bibinfo{author}{\bibfnamefont{A.}~\bibnamefont{Hubert}},
  \bibinfo{journal}{Phys. Stat. Sol. (b)} \textbf{\bibinfo{volume}{186}},
  \bibinfo{pages}{527} (\bibinfo{year}{1994}{\natexlab{b}}).

\bibitem[{\citenamefont{R{\"o}{\ss}ler
  et~al.}(2006)\citenamefont{R{\"o}{\ss}ler, Bogdanov, and
  Pfleiderer}}]{RBP_06}
\bibinfo{author}{\bibfnamefont{U.}~\bibnamefont{R{\"o}{\ss}ler}},
  \bibinfo{author}{\bibfnamefont{A.}~\bibnamefont{Bogdanov}}, \bibnamefont{and}
  \bibinfo{author}{\bibfnamefont{C.}~\bibnamefont{Pfleiderer}},
  \bibinfo{journal}{Nature} \textbf{\bibinfo{volume}{442}},
  \bibinfo{pages}{797} (\bibinfo{year}{2006}).

\bibitem[{\citenamefont{R{\"o}{\ss}ler
  et~al.}(2011)\citenamefont{R{\"o}{\ss}ler, Leonov, and Bogdanov}}]{ROSS_11}
\bibinfo{author}{\bibfnamefont{U.~K.} \bibnamefont{R{\"o}{\ss}ler}},
  \bibinfo{author}{\bibfnamefont{A.~A.} \bibnamefont{Leonov}},
  \bibnamefont{and} \bibinfo{author}{\bibfnamefont{A.~N.}
  \bibnamefont{Bogdanov}}, in \emph{\bibinfo{booktitle}{Journal of Physics:
  Conference Series}} (\bibinfo{organization}{IOP Publishing},
  \bibinfo{year}{2011}), vol. \bibinfo{volume}{303}, p.
  \bibinfo{pages}{012105}.

\bibitem[{\citenamefont{Leonov et~al.}(2016)\citenamefont{Leonov, Monchesky,
  Romming, Kubetzka, Bogdanov, and Wiesendanger}}]{LEO_16}
\bibinfo{author}{\bibfnamefont{A.}~\bibnamefont{Leonov}},
  \bibinfo{author}{\bibfnamefont{T.}~\bibnamefont{Monchesky}},
  \bibinfo{author}{\bibfnamefont{N.}~\bibnamefont{Romming}},
  \bibinfo{author}{\bibfnamefont{A.}~\bibnamefont{Kubetzka}},
  \bibinfo{author}{\bibfnamefont{A.}~\bibnamefont{Bogdanov}}, \bibnamefont{and}
  \bibinfo{author}{\bibfnamefont{R.}~\bibnamefont{Wiesendanger}},
  \bibinfo{journal}{New Journal of Physics} \textbf{\bibinfo{volume}{18}},
  \bibinfo{pages}{065003} (\bibinfo{year}{2016}).

\bibitem[{\citenamefont{Okamura et~al.}(2016)\citenamefont{Okamura, Kagawa,
  Seki, and Tokura}}]{OKAMURA_16}
\bibinfo{author}{\bibfnamefont{Y.}~\bibnamefont{Okamura}},
  \bibinfo{author}{\bibfnamefont{F.}~\bibnamefont{Kagawa}},
  \bibinfo{author}{\bibfnamefont{S.}~\bibnamefont{Seki}}, \bibnamefont{and}
  \bibinfo{author}{\bibfnamefont{Y.}~\bibnamefont{Tokura}},
  \bibinfo{journal}{Nature communications} \textbf{\bibinfo{volume}{7}},
  \bibinfo{pages}{12669} (\bibinfo{year}{2016}).

\bibitem[{\citenamefont{White et~al.}(2014)\citenamefont{White, Pr{\v{s}}a,
  Huang, Omrani, {\v{Z}}ivkovi{\'c}, Bartkowiak, Berger, Magrez, Gavilano, Nagy
  et~al.}}]{WHITE_14}
\bibinfo{author}{\bibfnamefont{J.}~\bibnamefont{White}},
  \bibinfo{author}{\bibfnamefont{K.}~\bibnamefont{Pr{\v{s}}a}},
  \bibinfo{author}{\bibfnamefont{P.}~\bibnamefont{Huang}},
  \bibinfo{author}{\bibfnamefont{A.}~\bibnamefont{Omrani}},
  \bibinfo{author}{\bibfnamefont{I.}~\bibnamefont{{\v{Z}}ivkovi{\'c}}},
  \bibinfo{author}{\bibfnamefont{M.}~\bibnamefont{Bartkowiak}},
  \bibinfo{author}{\bibfnamefont{H.}~\bibnamefont{Berger}},
  \bibinfo{author}{\bibfnamefont{A.}~\bibnamefont{Magrez}},
  \bibinfo{author}{\bibfnamefont{J.}~\bibnamefont{Gavilano}},
  \bibinfo{author}{\bibfnamefont{G.}~\bibnamefont{Nagy}}, \bibnamefont{et~al.},
  \bibinfo{journal}{Physical review letters} \textbf{\bibinfo{volume}{113}},
  \bibinfo{pages}{107203} (\bibinfo{year}{2014}).

\bibitem[{\citenamefont{Jiang et~al.}(2016)\citenamefont{Jiang, Zhang, Yu,
  Jungfleisch, Upadhyaya, Somaily, Pearson, Tserkovnyak, Wang, Heinonen
  et~al.}}]{JIA_16}
\bibinfo{author}{\bibfnamefont{W.}~\bibnamefont{Jiang}},
  \bibinfo{author}{\bibfnamefont{W.}~\bibnamefont{Zhang}},
  \bibinfo{author}{\bibfnamefont{G.}~\bibnamefont{Yu}},
  \bibinfo{author}{\bibfnamefont{M.~B.} \bibnamefont{Jungfleisch}},
  \bibinfo{author}{\bibfnamefont{P.}~\bibnamefont{Upadhyaya}},
  \bibinfo{author}{\bibfnamefont{H.}~\bibnamefont{Somaily}},
  \bibinfo{author}{\bibfnamefont{J.~E.} \bibnamefont{Pearson}},
  \bibinfo{author}{\bibfnamefont{Y.}~\bibnamefont{Tserkovnyak}},
  \bibinfo{author}{\bibfnamefont{K.~L.} \bibnamefont{Wang}},
  \bibinfo{author}{\bibfnamefont{O.}~\bibnamefont{Heinonen}},
  \bibnamefont{et~al.}, \bibinfo{journal}{AIP Advances}
  \textbf{\bibinfo{volume}{6}}, \bibinfo{pages}{055602} (\bibinfo{year}{2016}).

\bibitem[{\citenamefont{Woo et~al.}(2016)\citenamefont{Woo, Litzius,
  Kr{\"u}ger, Im, Caretta, Richter, Mann, Krone, Reeve, Weigand
  et~al.}}]{WOO_16}
\bibinfo{author}{\bibfnamefont{S.}~\bibnamefont{Woo}},
  \bibinfo{author}{\bibfnamefont{K.}~\bibnamefont{Litzius}},
  \bibinfo{author}{\bibfnamefont{B.}~\bibnamefont{Kr{\"u}ger}},
  \bibinfo{author}{\bibfnamefont{M.-Y.} \bibnamefont{Im}},
  \bibinfo{author}{\bibfnamefont{L.}~\bibnamefont{Caretta}},
  \bibinfo{author}{\bibfnamefont{K.}~\bibnamefont{Richter}},
  \bibinfo{author}{\bibfnamefont{M.}~\bibnamefont{Mann}},
  \bibinfo{author}{\bibfnamefont{A.}~\bibnamefont{Krone}},
  \bibinfo{author}{\bibfnamefont{R.~M.} \bibnamefont{Reeve}},
  \bibinfo{author}{\bibfnamefont{M.}~\bibnamefont{Weigand}},
  \bibnamefont{et~al.}, \bibinfo{journal}{Nature materials}
  \textbf{\bibinfo{volume}{15}}, \bibinfo{pages}{501} (\bibinfo{year}{2016}).

\bibitem[{\citenamefont{Thiele}(1969)}]{THI_69}
\bibinfo{author}{\bibfnamefont{A.~A.} \bibnamefont{Thiele}},
  \bibinfo{journal}{Bell Syst. Tech. J.} \textbf{\bibinfo{volume}{8}},
  \bibinfo{pages}{3287} (\bibinfo{year}{1969}).

\bibitem[{\citenamefont{Thiele}(1970)}]{THI_70}
\bibinfo{author}{\bibfnamefont{A.~A.} \bibnamefont{Thiele}},
  \bibinfo{journal}{J. Appl. Phys.} \textbf{\bibinfo{volume}{41}},
  \bibinfo{pages}{1139} (\bibinfo{year}{1970}).

\bibitem[{\citenamefont{Shibata et~al.}(2015)\citenamefont{Shibata, Iwasaki,
  Kanazawa, Aizawa, Tanigaki, Shirai, Nakajima, Kubota, Kawasaki, Park
  et~al.}}]{SHI_15}
\bibinfo{author}{\bibfnamefont{K.}~\bibnamefont{Shibata}},
  \bibinfo{author}{\bibfnamefont{J.}~\bibnamefont{Iwasaki}},
  \bibinfo{author}{\bibfnamefont{N.}~\bibnamefont{Kanazawa}},
  \bibinfo{author}{\bibfnamefont{S.}~\bibnamefont{Aizawa}},
  \bibinfo{author}{\bibfnamefont{T.}~\bibnamefont{Tanigaki}},
  \bibinfo{author}{\bibfnamefont{M.}~\bibnamefont{Shirai}},
  \bibinfo{author}{\bibfnamefont{T.}~\bibnamefont{Nakajima}},
  \bibinfo{author}{\bibfnamefont{M.}~\bibnamefont{Kubota}},
  \bibinfo{author}{\bibfnamefont{M.}~\bibnamefont{Kawasaki}},
  \bibinfo{author}{\bibfnamefont{H.}~\bibnamefont{Park}}, \bibnamefont{et~al.},
  \bibinfo{journal}{Nature nanotechnology} \textbf{\bibinfo{volume}{10}},
  \bibinfo{pages}{589} (\bibinfo{year}{2015}).

\bibitem[{\citenamefont{Makhfudz et~al.}(2012)\citenamefont{Makhfudz,
  Kr{\"u}ger, and Tchernyshyov}}]{TCH_12}
\bibinfo{author}{\bibfnamefont{I.}~\bibnamefont{Makhfudz}},
  \bibinfo{author}{\bibfnamefont{B.}~\bibnamefont{Kr{\"u}ger}},
  \bibnamefont{and}
  \bibinfo{author}{\bibfnamefont{O.}~\bibnamefont{Tchernyshyov}},
  \bibinfo{journal}{Physical review letters} \textbf{\bibinfo{volume}{109}},
  \bibinfo{pages}{217201} (\bibinfo{year}{2012}).

\bibitem[{\citenamefont{Lin and Saxena}(2015)}]{LIN_15}
\bibinfo{author}{\bibfnamefont{S.-Z.} \bibnamefont{Lin}} \bibnamefont{and}
  \bibinfo{author}{\bibfnamefont{A.}~\bibnamefont{Saxena}},
  \bibinfo{journal}{Physical Review B} \textbf{\bibinfo{volume}{92}},
  \bibinfo{pages}{180401} (\bibinfo{year}{2015}).

\bibitem[{\citenamefont{Mochizuki and Seki}(2013)}]{MOCHI_13}
\bibinfo{author}{\bibfnamefont{M.}~\bibnamefont{Mochizuki}} \bibnamefont{and}
  \bibinfo{author}{\bibfnamefont{S.}~\bibnamefont{Seki}},
  \bibinfo{journal}{Physical Review B} \textbf{\bibinfo{volume}{87}},
  \bibinfo{pages}{134403} (\bibinfo{year}{2013}).

\bibitem[{\citenamefont{Mochizuki and Watanabe}(2015)}]{MOCHI_15}
\bibinfo{author}{\bibfnamefont{M.}~\bibnamefont{Mochizuki}} \bibnamefont{and}
  \bibinfo{author}{\bibfnamefont{Y.}~\bibnamefont{Watanabe}},
  \bibinfo{journal}{Applied Physics Letters} \textbf{\bibinfo{volume}{107}},
  \bibinfo{pages}{082409} (\bibinfo{year}{2015}).

\bibitem[{\citenamefont{Mochizuki}(2016)}]{MOCHI_16}
\bibinfo{author}{\bibfnamefont{M.}~\bibnamefont{Mochizuki}},
  \bibinfo{journal}{Advanced Electronic Materials}
  \textbf{\bibinfo{volume}{2}}, \bibinfo{pages}{1500180}
  (\bibinfo{year}{2016}).

\bibitem[{\citenamefont{Hubert and Sch{\"a}fer}(1998)}]{HUB_SCH_98}
\bibinfo{author}{\bibfnamefont{A.}~\bibnamefont{Hubert}} \bibnamefont{and}
  \bibinfo{author}{\bibfnamefont{R.}~\bibnamefont{Sch{\"a}fer}},
  \emph{\bibinfo{title}{Magnetic domains: the analysis of magnetic
  microstructures}} (\bibinfo{year}{1998}).

\bibitem[{\citenamefont{Bran et~al.}(2009)\citenamefont{Bran, Butenko, Kiselev,
  Wolff, Schultz, Hellwig, R{\"o}{\ss}ler, Bogdanov, and Neu}}]{BRAN_09}
\bibinfo{author}{\bibfnamefont{C.}~\bibnamefont{Bran}},
  \bibinfo{author}{\bibfnamefont{A.~B.} \bibnamefont{Butenko}},
  \bibinfo{author}{\bibfnamefont{N.~S.} \bibnamefont{Kiselev}},
  \bibinfo{author}{\bibfnamefont{U.}~\bibnamefont{Wolff}},
  \bibinfo{author}{\bibfnamefont{L.}~\bibnamefont{Schultz}},
  \bibinfo{author}{\bibfnamefont{O.}~\bibnamefont{Hellwig}},
  \bibinfo{author}{\bibfnamefont{U.~K.} \bibnamefont{R{\"o}{\ss}ler}},
  \bibinfo{author}{\bibfnamefont{A.~N.} \bibnamefont{Bogdanov}},
  \bibnamefont{and} \bibinfo{author}{\bibfnamefont{V.}~\bibnamefont{Neu}},
  \bibinfo{journal}{Physical Review B} \textbf{\bibinfo{volume}{79}},
  \bibinfo{pages}{024430} (\bibinfo{year}{2009}).

\bibitem[{\citenamefont{Han et~al.}(2010)\citenamefont{Han, Zang, Yang, Park,
  and Nagaosa}}]{HZY_10}
\bibinfo{author}{\bibfnamefont{J.~H.} \bibnamefont{Han}},
  \bibinfo{author}{\bibfnamefont{J.}~\bibnamefont{Zang}},
  \bibinfo{author}{\bibfnamefont{Z.}~\bibnamefont{Yang}},
  \bibinfo{author}{\bibfnamefont{J.-H.} \bibnamefont{Park}}, \bibnamefont{and}
  \bibinfo{author}{\bibfnamefont{N.}~\bibnamefont{Nagaosa}},
  \bibinfo{journal}{Physical Review B} \textbf{\bibinfo{volume}{82}},
  \bibinfo{pages}{094429} (\bibinfo{year}{2010}).

\bibitem[{\citenamefont{Wang et~al.}(2017)\citenamefont{Wang, Du, Zhao, Jin,
  Tian, Zhang, and Che}}]{WAN_17}
\bibinfo{author}{\bibfnamefont{C.}~\bibnamefont{Wang}},
  \bibinfo{author}{\bibfnamefont{H.}~\bibnamefont{Du}},
  \bibinfo{author}{\bibfnamefont{X.}~\bibnamefont{Zhao}},
  \bibinfo{author}{\bibfnamefont{C.}~\bibnamefont{Jin}},
  \bibinfo{author}{\bibfnamefont{M.}~\bibnamefont{Tian}},
  \bibinfo{author}{\bibfnamefont{Y.}~\bibnamefont{Zhang}}, \bibnamefont{and}
  \bibinfo{author}{\bibfnamefont{R.}~\bibnamefont{Che}}, \bibinfo{journal}{Nano
  letters} \textbf{\bibinfo{volume}{17}}, \bibinfo{pages}{2921}
  (\bibinfo{year}{2017}).

\bibitem[{\citenamefont{White et~al.}(2012)\citenamefont{White, Levatic,
  Omrani, Egetenmeyer, Prsa, Zivkovic, Gavilano, Kohlbrecher, Bartkowiak,
  Berger et~al.}}]{JSWHITE_12}
\bibinfo{author}{\bibfnamefont{J.~S.} \bibnamefont{White}},
  \bibinfo{author}{\bibfnamefont{I.}~\bibnamefont{Levatic}},
  \bibinfo{author}{\bibfnamefont{A.~A.} \bibnamefont{Omrani}},
  \bibinfo{author}{\bibfnamefont{N.}~\bibnamefont{Egetenmeyer}},
  \bibinfo{author}{\bibfnamefont{K.}~\bibnamefont{Prsa}},
  \bibinfo{author}{\bibfnamefont{I.}~\bibnamefont{Zivkovic}},
  \bibinfo{author}{\bibfnamefont{J.~L.} \bibnamefont{Gavilano}},
  \bibinfo{author}{\bibfnamefont{J.}~\bibnamefont{Kohlbrecher}},
  \bibinfo{author}{\bibfnamefont{M.}~\bibnamefont{Bartkowiak}},
  \bibinfo{author}{\bibfnamefont{H.}~\bibnamefont{Berger}},
  \bibnamefont{et~al.}, \bibinfo{journal}{Journal of Physics: Condensed Matter}
  \textbf{\bibinfo{volume}{24}}, \bibinfo{pages}{432201}
  (\bibinfo{year}{2012}).

\bibitem[{\citenamefont{Mochizuki}(2017)}]{MOCH_16}
\bibinfo{author}{\bibfnamefont{M.}~\bibnamefont{Mochizuki}},
  \bibinfo{journal}{Applied Physics Letters} \textbf{\bibinfo{volume}{111}},
  \bibinfo{pages}{092403} (\bibinfo{year}{2017}).

\end{thebibliography}

\end{document}